\documentclass{osa-article}
\usepackage{graphicx}
\usepackage{amsmath,amsfonts}
\usepackage{xcolor, fontawesome}
\usepackage[nolist]{acronym}
\usepackage{tikz}
\usepackage[ruled,vlined]{algorithm2e}
\usepackage{xspace}

\def\toliman{\textsc{Toliman}\xspace} 
\definecolor{linkcolor}{rgb}{0.1216,0.4667,0.7059}
\newacro{mse}[MSE]{Mean Squared Error}
\newacro{mae}[MAE]{Mean Absolute Error} 
\newacro{app}[APP]{Apodizing Phase Plate}

\journal{osac}

\articletype{Research Article}

\begin{document}

\title{Phase Retrieval and Design with Automatic Differentiation}

\author{Alison Wong,\authormark{1,*} Benjamin Pope\authormark{2,3},  Louis Desdoigts\authormark{1}, Peter Tuthill\authormark{1}, Barnaby Norris\authormark{1}, Chris Betters\authormark{1}}

\address{\authormark{1}Sydney Institute for Astronomy (SIfA), School of Physics, The University of Sydney, NSW 2006, Australia\\
\authormark{2}School of Mathematics and Physics, The University of Queensland, St Lucia, QLD 4072, Australia\\
\authormark{3}Centre for Astrophysics, University of Southern Queensland, West Street, Toowoomba, QLD 4350, Australia
}

\email{\authormark{*}a.wong@sydney.edu.au} 



\begin{abstract}
The principal limitation in many areas of astronomy, especially for directly imaging exoplanets, arises from instability in the point spread function (PSF) delivered by the telescope and instrument. 
To understand the transfer function, it is often necessary to infer a set of optical aberrations given only the intensity distribution on the sensor --- the problem of {\it phase retrieval}.
This can be important for post-processing of existing data, or for the design of optical phase masks to engineer PSFs optimized to achieve high contrast, angular resolution, or astrometric stability. 
By exploiting newly efficient and flexible technology for {\it automatic differentiation,} which in recent years has undergone rapid development driven by machine learning, we can perform both phase retrieval and design in a way that is systematic, user-friendly, fast, and effective.
By using modern gradient descent techniques, this converges efficiently and is easily extended to incorporate constraints and regularization.
We illustrate the wide-ranging potential for this approach using our new package, \textsc{Morphine}.
Challenging applications performed with this code include precise phase retrieval for both discrete and continuous phase distributions, even where information has been censored such as heavily-saturated sensor data. 
We also show that the same algorithms can optimize continuous or binary phase masks that are competitive with existing best solutions for two example problems: an \ac{app} coronagraph for exoplanet direct imaging, and a diffractive pupil for narrow-angle astrometry.
The \textsc{Morphine} source code and examples are available open-source, with a similar interface to the popular physical optics package \textsc{Poppy}. 
\href{https://github.com/alipwong/phase_retrieval_and_design}{\color{linkcolor}\faGithub}
\end{abstract}


\section{Introduction}

\subsection{Phase retrieval}


Phase retrieval is the task of recovering phase information from a PSF and is relevant to a range of optical science and engineering including electron microscopy, crystallography, coherent diffractive imaging and astronomy. Famously in astronomy, phase retrieval was applied to precisely measure and correct for the \textit{Hubble Space Telescope} mirror aberration shortly after its launch \cite{fienup93}. Correcting for phase aberrations is essential in pushing space telescopes to achieve the high angular resolution and contrast necessary to detect planets around other stars \cite{ygouf2013}.

The phase retrieval problem arises from the inability of detectors (i.e. cameras) to measure the phase of incoming radiation directly. The PSF is the far-field diffraction pattern of the telescope, which at focus (for ideal optics) is the Fourier Transform of the electric field distribution in the pupil plane. Unlike measurements made at radio wavelengths where electric fields can be measured directly, optical detectors measure an intensity distribution given by the squared modulus of the complex PSF convolved with the source intensity distribution. This measurement suffers a loss of information at the detector: taking the square modulus eliminates phase so that the original electric field cannot be recovered by reversing the mathematical operations using an inverse Fourier transform. 


Phase retrieval is an intrinsically difficult problem as for geometric reasons it is not well-posed, however in practice a number of adequate solutions have been found \cite{barnett2020}. Two early methods still used today are the Gerchberg-Saxton (GS) algorithm \cite{gerchberg1972} and the hybrid input-output (HIO) algorithm \cite{fienup1978}. Both are iterative methods that start with an initial phase estimate and then alternate propagation between the focal and pupil planes, each time imposing constraints so that the solution conforms to known information. Further work has seen the development of the HIO algorithm to incorporate sparsity constraints \cite{Mukherjee2012}.
A comparison of these iterative methods (GS and HIO) is given in \cite{fienup1982}.

Numerous other approaches have been taken to solving phase retrieval owing to its central importance in so many areas of science. 
Although a more complete review of the extensive literature in this area is beyond the scope here, the interested reader is directed to matrix completion \cite{candes2011} and Wirtinger flow \cite{candes2014}; semi-definite programming with sparsity constraints \cite{Shechtman2011,ohlsson2012,waldspurger2015}; or greedy algorithms with sparsity constraints \cite{bahmani2011,schechtman2014}; as summarized in \cite{schechtman2015}. Deep learning has proven successful in phase retrieval \cite{kappeler2017,boominathan2018,metzler2018,jagatap2019,wang2020,nishizaki2020}, and while we will use the underlying technology of automatic differentiation in the present work, we focus exclusively on using deterministic optical simulation.

\subsection{Automatic Differentiation}

Gradient-based search algorithms provide another strategy that can be applied to the phase retrieval problem. While straightforward gradient descent can perform poorly on nonlinear problems, becoming trapped in local minima, it is often the only practical method for optimizing over a high-dimensional parameter space such as in neural networks, or large and complex phase patterns. Automatic differentiation, or autodiff, has made the well known gradient based `backpropagation' algorithm \cite{lecun1988theoretical} newly efficient, and is a key enabling technolgy in the burgeoning field of deep learning \cite{lecun15,baydin2015}. 

Autodiff is extremely well suited to optical simulation and optimization. A neural network consists of a series of alternating linear transformations of a vector, generally followed by nonlinear element-wise operations.
This is isomorphic to the linear propagation (Fourier or Fresnel) and local phase and amplitude masking in optical simulation. 

Interestingly, the analogy also runs in reverse: it is possible to implement a neural network for image classification entirely using optical phase and amplitude masks, so that light (for example) from an image of a handwritten digit or fashion product will diffract onto a spot corresponding to this classification \cite{lin2018}. This deep isomorphism means that the same software libraries and optimization approaches that are efficient for neural network implementations can be straightforwardly applied to optical simulation.

Autodiff allows the direct integration of deterministic optical simulation with neural networks. This is distinct from its use in neural networks, which has been the most common application in the field so far, and instead we are interested in using it for the physical optics simulation itself. Autodiff has been applied to geometric optics \cite{sutin2016}, including gravitational lensing \cite{morningstar2019,karchev2021}, and to interferometric imaging \cite{czekala2020}. 

A framework was established by Jurling \& Fienup \cite{jurling2014} applying algorithmic differentiation to phase retrieval, in which the differentiation rules for common operations were obtained by hand. One of the major themes of the present paper is to build upon this idea applying modern automatic differentiation libraries, and flexibly linking it with practical, standard optical models. This can then be used for data analysis tasks, and extending this phase retrieval idea to hardware design. Because of the popularity of neural networks, there has recently been a very considerable public and private investment in both software and hardware for autodiff, 
which we can here exploit for physical optics. There are many Python implementations of autodiff, including \textsc{theano} \cite{theano}, \textsc{PyTorch} \cite{pytorch}, \textsc{TensorFlow} \cite{tensorflow2015}, \textsc{autograd} \cite{autograd}, and \textsc{Jax} \cite{jax} which closely follows the \textsc{NumPY} API \cite{numpy}. Similarly, the interpreted language Julia has native support for autodiff \cite{julia}. These libraries offer much greater flexibility for phase retrieval than was possible in the original framework \cite{jurling2014}, with work done in Molnar \& Nikolic \cite{MolnarNikolic2020} demonstrating that \textsc{Jax} can be applied to phase retrieval problem in radio astronomy.




Gradient-based techniques only require a differentiable objective function and so, can be trivially applied to a wide variety of tasks. In this paper, we look at applying these techniques to challenges encountered in optics, which are naturally primed for this framework as the underlying operations are straightforwardly differentiable. These tasks are therefore very well-suited to gradient descent optimizations or even deep learning solutions. With regularization terms, or with a thresholded parametrization, we show that it is straightforward to optimize not just continuous, but also discrete phase masks. We will address example problems in both phase retrieval and phase design in the context of astronomy. 

In the first half of this paper we apply similar methods of phase retrieval as \cite{wang2020}, but in the context of optical astronomy. Phase retrieval in astronomy presents its own set of unique challenges, such as nonlinearities introduced by detector saturation. 

We demonstrate phase retrieval with gradient descent, including the effects of saturation. We will take the saturation model and linear detector sensitivity map as known, but note that these can be parametrized, differentiated, and learned as part of the same procedure (Desdoigts et al, in prep.) We also demonstrate how constraints can be added into the retrieval process, which allows for the incorporation of domain knowledge. The advantages in phase retrieval provided by having exact derivatives enable us to straightforwardly generalize to phase optic design, enabling fast gradient-based optimization of optical masks with respect to arbitrary objective functions.

\subsection{Designing Astronomical Phase Optics}


Astronomy requires high precision and low noise observations, relying heavily on the performance of optical instrumentation. Innovative design coupled with painstaking construction is therefore necessary to provide these quality measurements. We draw a distinction between phase retrieval, where phase screen consistent with a PSF is recovered, and phase design, where a phase mask is optimised to achieve a PSF maximizing some figure of merit. Phase design can be used to develop application-specific phase masks, to optimize some component of a much larger optical system. 

The challenge of phase design is related to that of phase retrieval and much of the machinery used in phase retrieval can be re-purposed for this task. The key adjustment is that optimization is performed with respect to some carefully constructed objective function instead of optimizing on a metric measuring the difference between the observed and estimated PSFs. The difficulty often lies in the balancing of various components in the objective function.

Phase design using a differentiable physical optics simulation has been successfully demonstrated in other areas of optics: for example, in adaptive optics \cite{vishniakou2020}; the \textsc{DeepOptics} project, optimizing optical components in cameras in the areas of super-resolution imaging and in extending depth of field \cite{sitzmann2018}; or \textsc{WaveBlocks} in microscopy \cite{page2020}. In each of these cases, a physical optics simulation is implemented in an autodiff framework and used to design phase optics by gradient descent. We will be applying a related approach for the first time to astronomical optics.

In this paper we demonstrate phase design on two test problems. The first is in the design of an \ac{app} coronagraph, and the second is in the design of a diffractive pupil for the \toliman space mission.

Coronagraphs are essential to high-contrast imaging and are prominently used in exoplanetary imaging. The goal of a coronagraph is to suppress on-axis starlight for some regions (or all) of the field of interest so that a fainter target, such as a planet, may be more easily identified. Among the most common and simplest forms of coronagraph design are those which result in a dark hole, such as an annular dark zone, within the PSF. These are characterized by the inner and outer working angles and evaluated by their contrast ratio. Due to their symmetry, these are generally easier to optimize compared to alternatives such as those enforcing a D-shaped region or a rectangular dark region.

An \ac{app} pupil plane coronagraph enforces a dark hole in the PSF purely by inducing a phase change \cite{codona2006} in the pupil plane. This results in full throughput of all starlight, and is distinct from a large family of alternative technologies in which light encounters obstructions in the pupil and/or focal plane. \ac{app} coronagraphs can be practically implemented with vector-Apodizing Phase Plates \cite{snik2012, otten2014a, otten2014b, otten2017, bos2021}, which apply liquid crystal technologies to manufacture intricate phase patterns.

APP coronagraphs redistribute unwanted starlight rather than remove it and despite their simplicity, have proven popular. They have been widely used on large ground-based telescopes world-wide, since the original tests at the MMT Observatory 6.5m telescope \cite{kenworthy2007}. They have since been used to directly image the exoplanets $\beta$~Pic\,b \cite{quanz2010}, HD~100546~b \cite{quanz2013}, and many other planets and substellar objects \cite{doelman2021}.

A less familiar application for phase design, re-examined here, is the diffractive pupil for the \toliman space telescope \cite{tuthill2018}. This is an ambitious mission to detect exoplanets in the $\alpha$ Cen A/B binary system by way of very high precision relative astrometry between $\alpha$ Cen A and $\alpha$ Cen B. In order to reach the precision required for exoplanetary detection, the \toliman mission takes the novel approach of using a diffractive pupil to create a richly-featured PSF. Observations of the binary system will then result in overlapping fringes, which can be used to make the required precise relative positional measurements.

\begin{figure}
    \centering
    \begin{tikzpicture}[scale = 1]
    	\node[inner sep=0pt] () at (0, -2) {\includegraphics[height = 3.5cm]{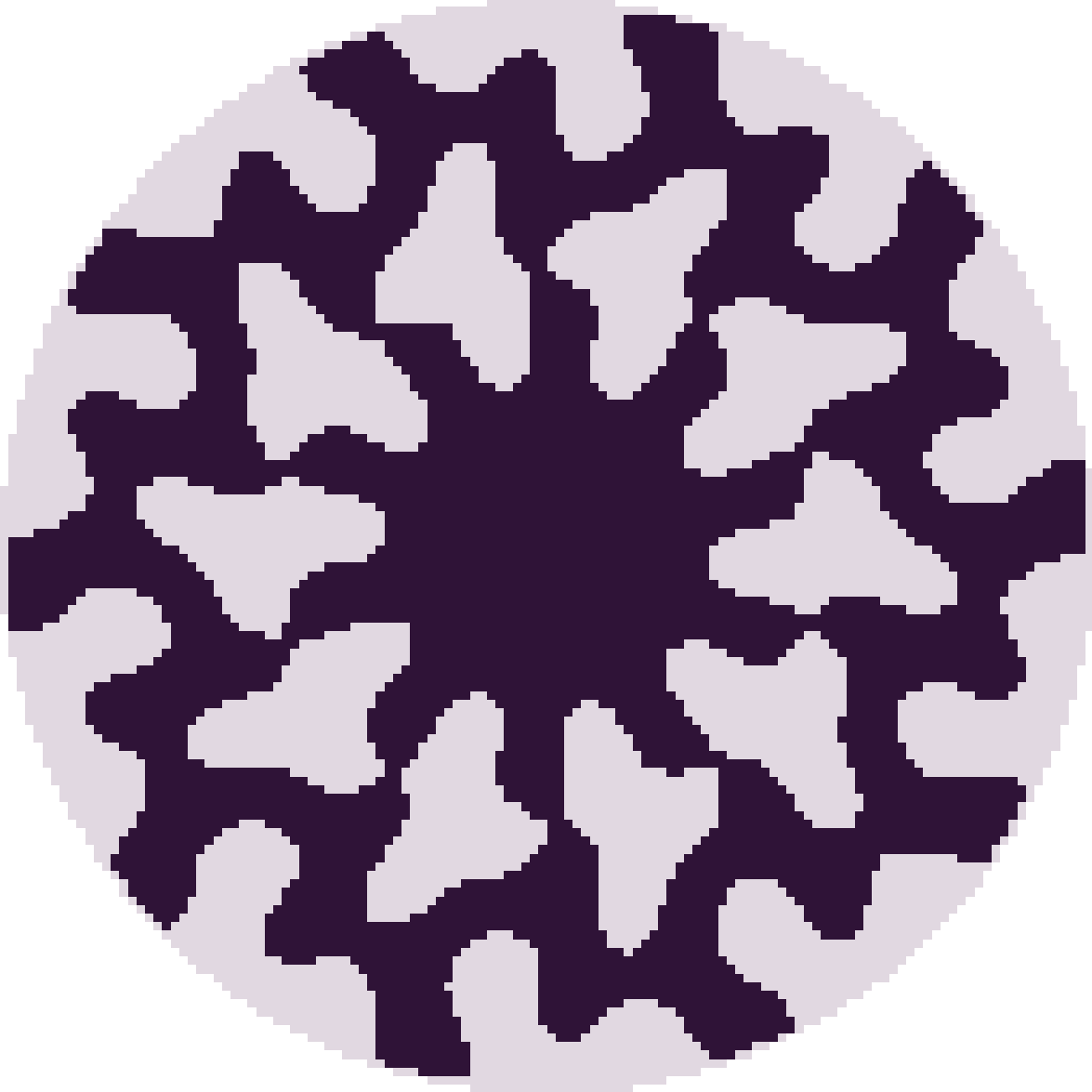}};
    
    	\node[inner sep=0pt] () at (2.5, -2.05) {\includegraphics[height = 4.05cm]{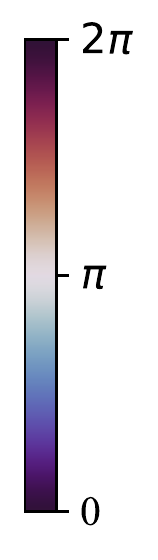}};
    	
    	\node[inner sep=0pt] () at (5, -2) {\includegraphics[width = 3.5cm]{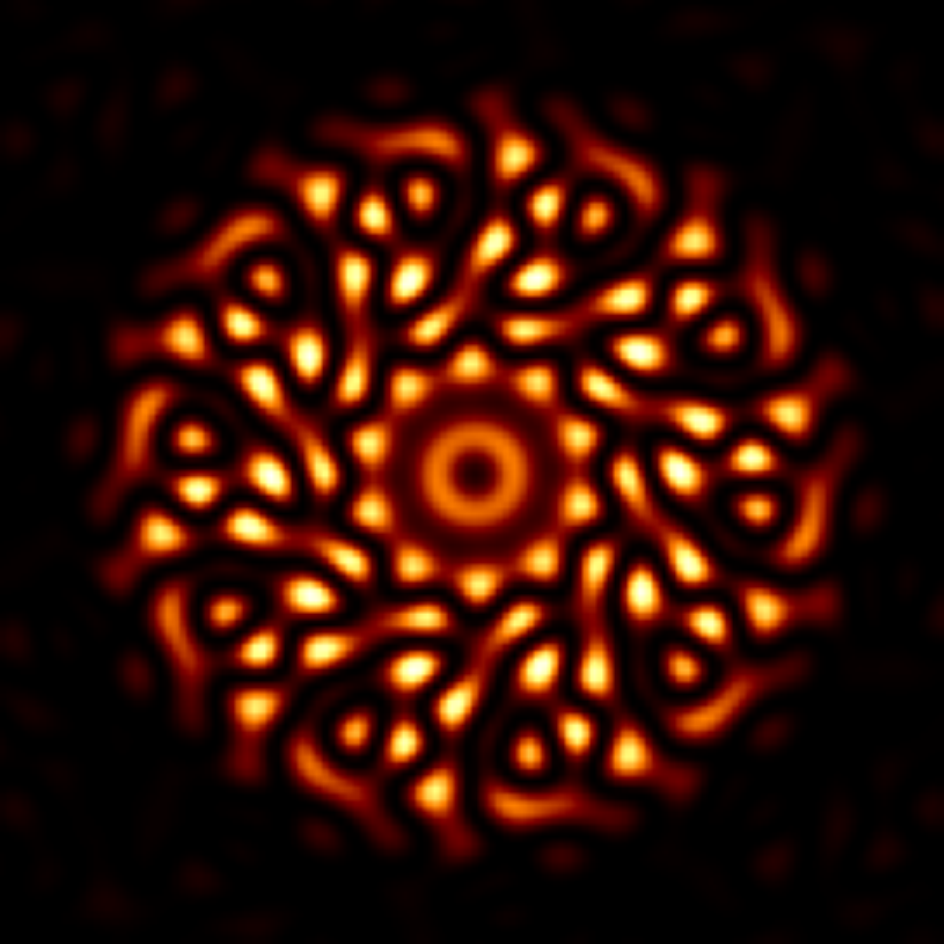}};
    
        \node[inner sep=0pt] () at (7.3, -2.04) {\includegraphics[height = 3.75cm]{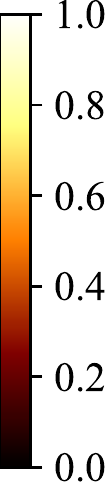}};
        
        \node[inner sep=0pt] () at (10, -2) {\includegraphics[height = 3.5cm]{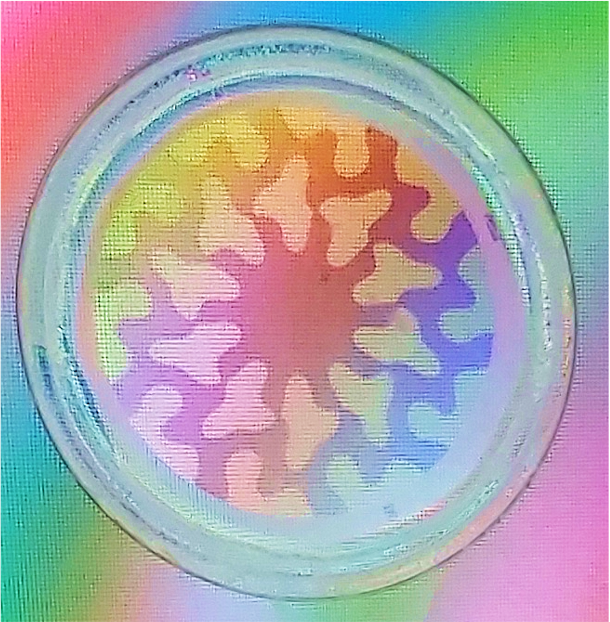}};
    
    \end{tikzpicture}
    \caption{Left: TinyTol diffractive pupil. Center: corresponding normalized PSF generated through simulation. Right: Photograph of the fabricated TinyTol pupil through two polarizing filters in order to display the phase pattern. \label{fig:true_toliman}}
    
\end{figure}

A candidate pupil for the \toliman mission and its corresponding PSF is shown in Figure~\ref{fig:true_toliman} and will be used in the TinyTol pathfinder mission, due to launch in August 2021. This will be a payload in the first CubeSat to be launched by CUAVA\footnote{The Australian Research Council Training Center for CubeSats, UAVs, and their Applications}, ISTI (the Image, Spectograph and TinyTol Instrument). TinyTol has an 18mm F/8.3 aperture and is a pathfinder/demonstrator for the larger \toliman telescope, which is expected to have a $\sim10$cm aperture.

The TinyTol pupil is composed of discrete regions with a $\pi$ phase step difference \cite{tuthill2018}. By construction, there are equal areas of 0 and $\pi$ phases, which induces a null at the origin of the PSF. This aids in diverting light into the fringes of the pattern and also compensates for light leakage, which will manifest as a central peak.

The final design requirements for the \toliman pupil are subject to onging review, but for the purposes of this paper there are two other major design considerations. Firstly, the PSF must span a large enough region of the detector, being uniformly filled with sharp features, so that the fringes from each star in the $\alpha$ Cen system overlap. This allows for the relative positional measurement to be made with respect to those fringes. Secondly, the pattern ascribed to the pupil must be made of reasonably sized contiguous regions in order to ensure the design is within realistic fabrication limits.


\section{Phase Retrieval}
\subsection{Methods}
Here we demonstrate phase retrieval on test problems to illustrate practical challenges encountered in astronomical phase retrieval.

\subsubsection{Pupil recovery by gradient descent}
As proof of concept, we perform phase retrieval from a target PSF with gradient descent for 3 cases:
\begin{enumerate}\itemsep 0em
	\item TinyTol: the phase solution is known (Figure~\ref{fig:true_toliman}).
	\item Quantum Monodromy (QM) pattern \cite{QM_pattern}: maintains a uniform density of points as a function of radius. Each of the peaks are constructed from a Gaussian with equal height and width, approximately the same size and shape of speckles. These properties were determined to be desirable for the \toliman PSF. This serves as an example of a realistic target PSF with desired properties, but where the phase solution is unknown and only an approximate solution is assumed to exist. 
	\item Faces of two coauthors: an arbitrary scene standing in for the PSF; here the phase solution is unknown and may not exist.
\end{enumerate}

\begin{figure}
    \centering
    \begin{tikzpicture}[scale = 1]
    
    \node at (-3, -2) [rotate = 90] {{Target PSF ($T$)}};
    \node at (-2.3, -2) [rotate = 90] {{Normalized: $\left(\frac{T}{\max(T)}\right)$}};
    \node at (-3, -6) [rotate = 90] {{Reconstructed PSF $(R)$}};
    \node at (-2.3, -6) [rotate = 90] {{Normalized:  $\left(\frac{R}{\max(T)}\right)$}};
    \node at (-3, -10) [rotate = 90] {{Difference Image}};
    \node at (-2.3, -10) [rotate = 90] {{$\left(\frac{|R - T|}{\max(T)}\right)$}};
    \node at (-2.3, -14) [rotate = 90] {{Recovered Pupil}};
    
    \node at (0.1, 0.2) [] {{TinyTol}};
    \node at (4.6, 0.55) [] {{Quantum Monodromy}};
    \node at (4.6, 0.2) [] {{Pattern}};
    \node at (9.5, 0.2) [] {{Faces}};
    

    \node[inner sep=0pt] () at (0, -2) {\includegraphics[width = 3.5cm]{figures/gradient_desc/toliman_true_psf_zoom_nocb.pdf}};
    \node[inner sep=0pt] () at (2.27, -2.05) {\includegraphics[height = 3.75cm]{figures/gradient_desc/toliman_true_psf_zoom_cb.pdf}};
    
    \node[inner sep=0pt] () at (0, -6) {\includegraphics[width = 3.5cm]{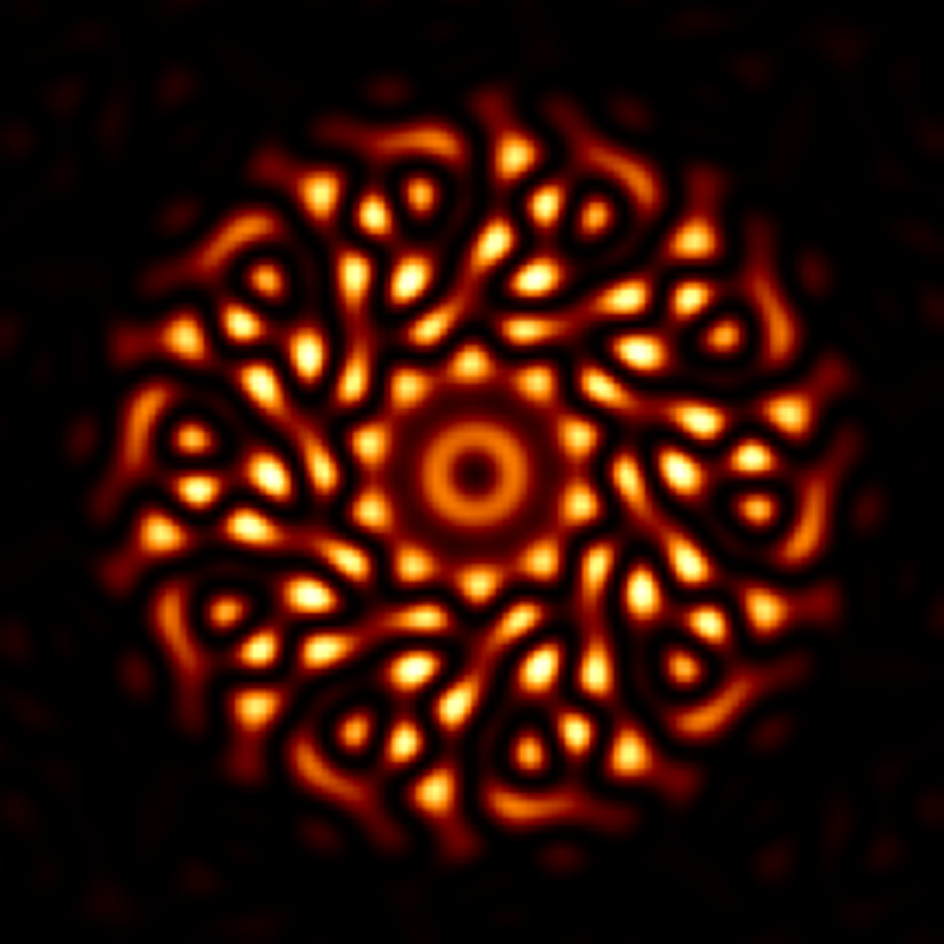}};
    \node[inner sep=0pt] () at (2.27, -6.05) {\includegraphics[height = 3.75cm]{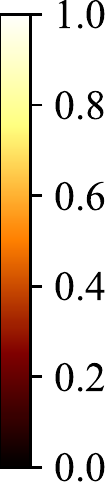}};
    
    \node[inner sep=0pt] () at (0, -10) {\includegraphics[width = 3.5cm]{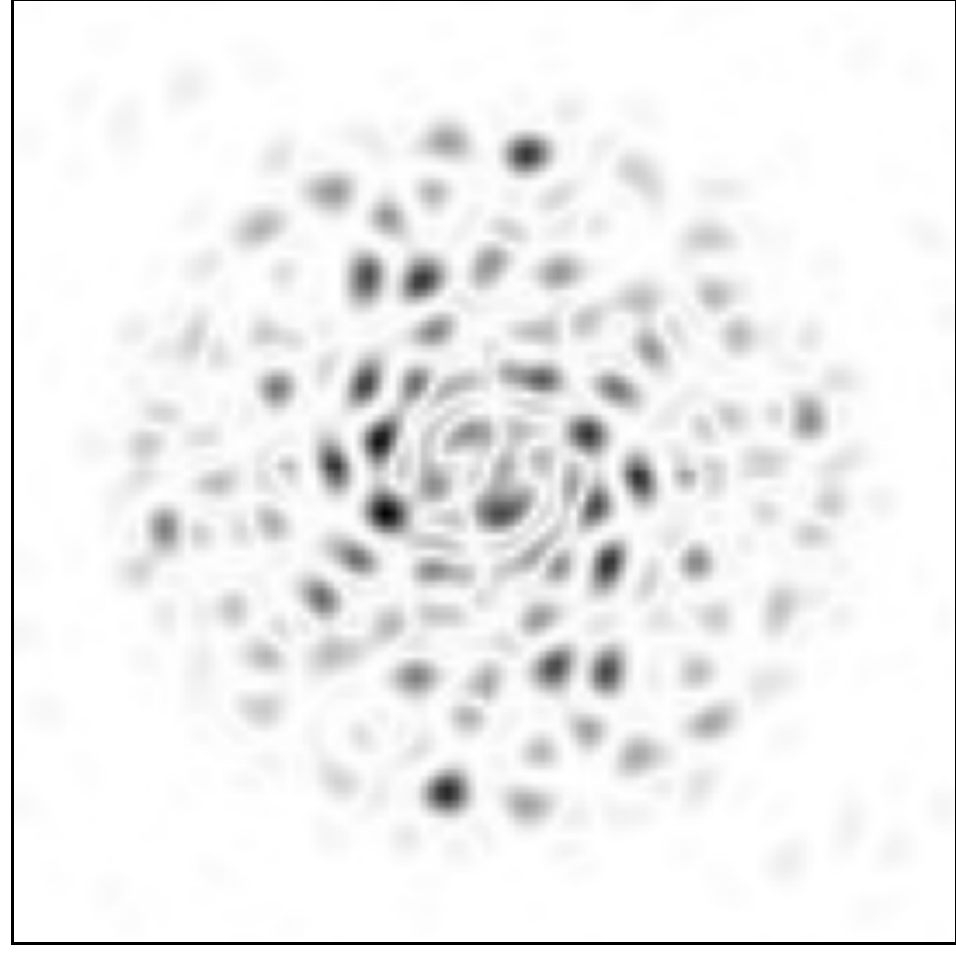}};
    \node[inner sep=0pt] () at (2.25, -9.9) {\includegraphics[height = 3.95cm]{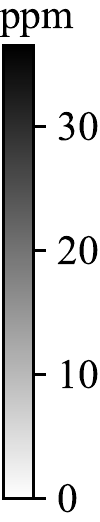}};
    
    \node[inner sep=0pt] () at (0, -13.85) {\includegraphics[width = 3.3cm]{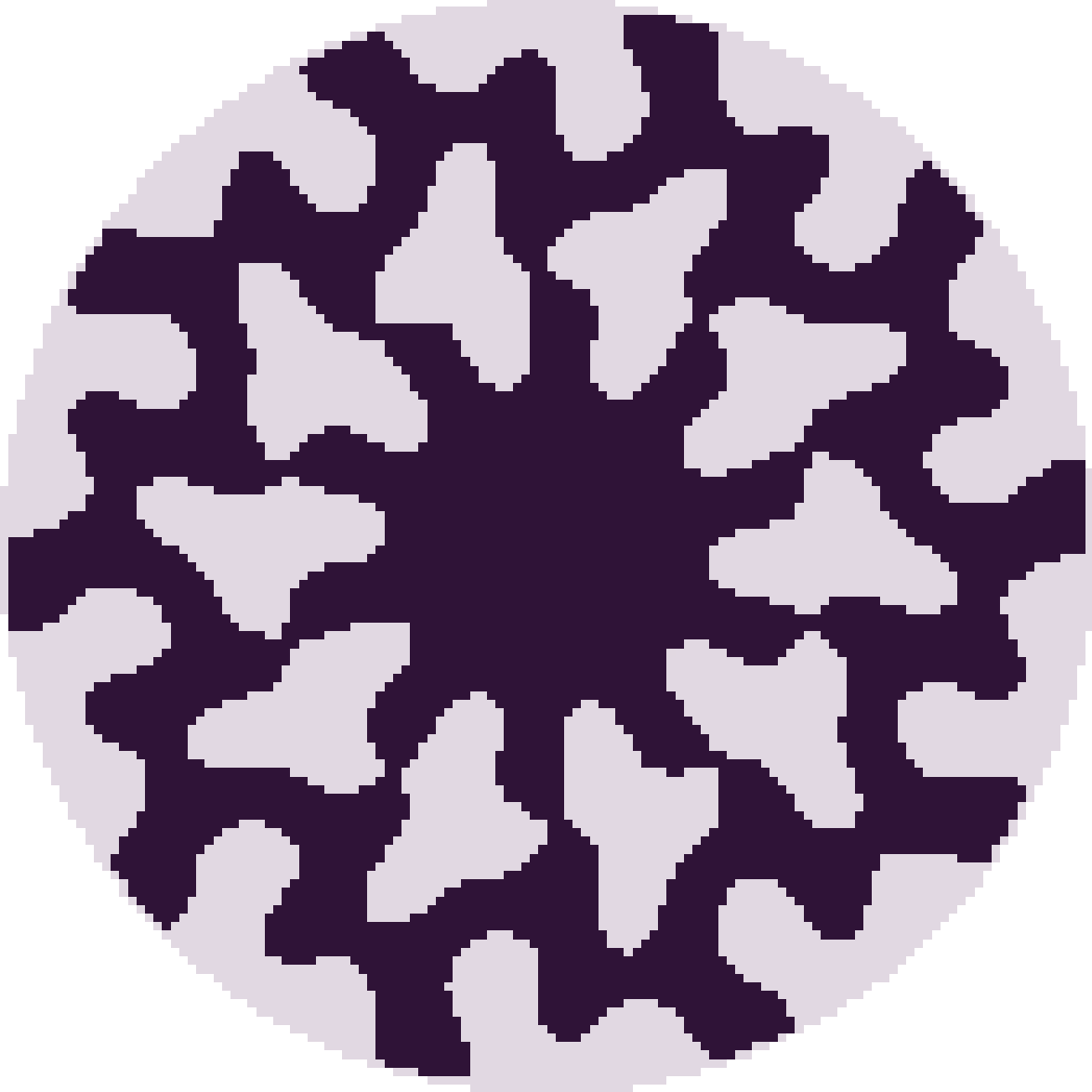}};
    \node[inner sep=0pt] () at (2.28, -13.85) {\includegraphics[height = 3.5cm]{figures/gradient_desc/toliman_true_pupil_cb.pdf}};
    
    \node[inner sep=0pt] () at (4.5, -2) {\includegraphics[width = 3.5cm]{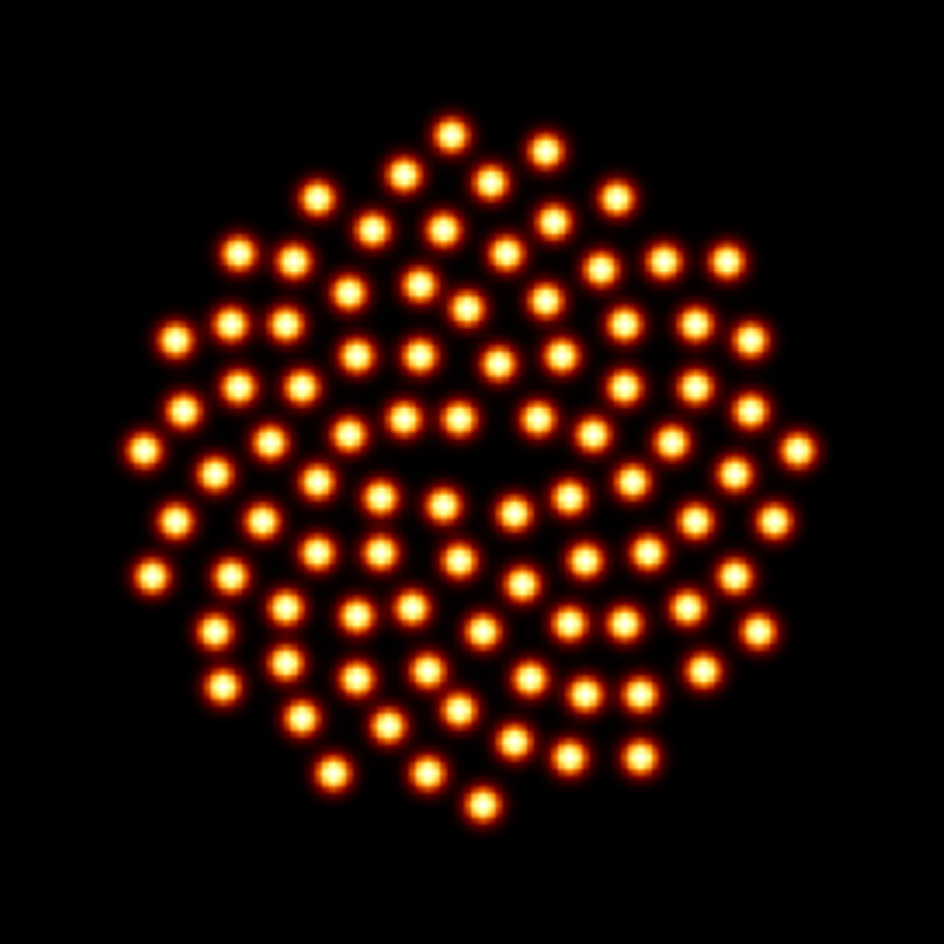}};
    \node[inner sep=0pt] () at (6.77, -2.05) {\includegraphics[height = 3.75cm]{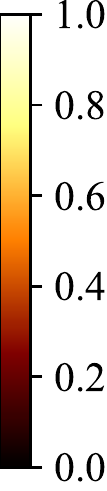}};
    
    \node[inner sep=0pt] () at (4.5, -6) {\includegraphics[width = 3.5cm]{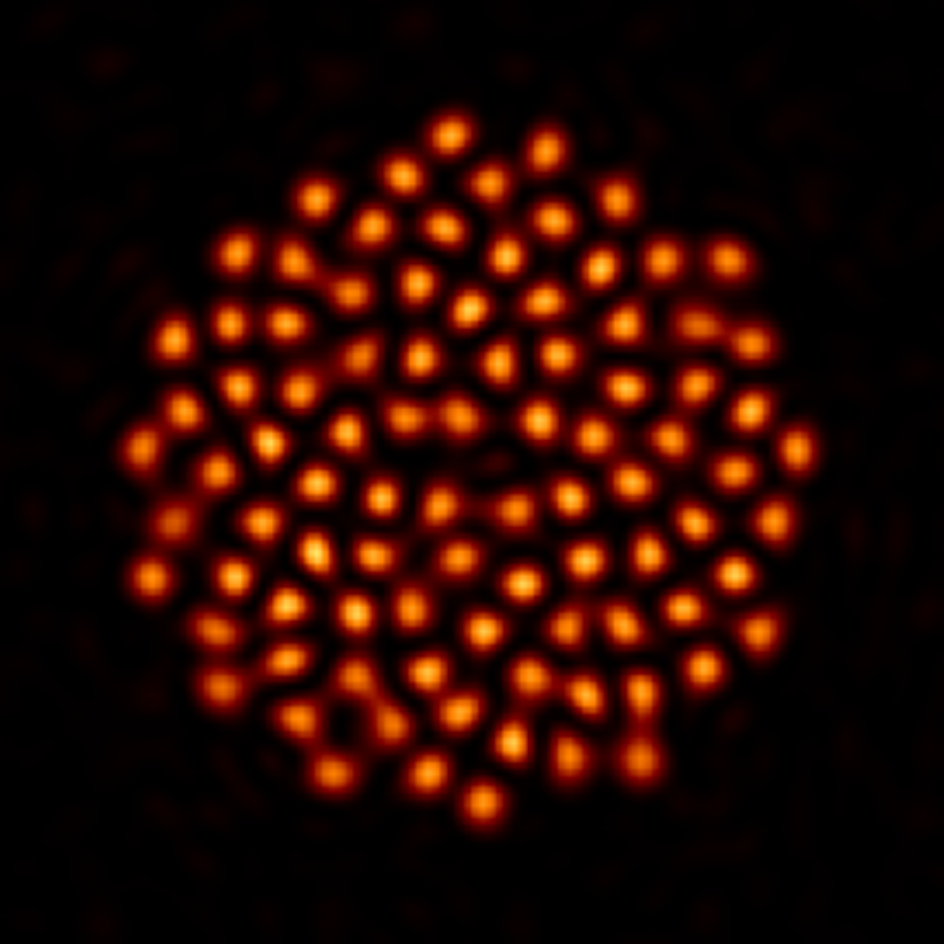}};
    \node[inner sep=0pt] () at (6.77, -6.05) {\includegraphics[height = 3.75cm]{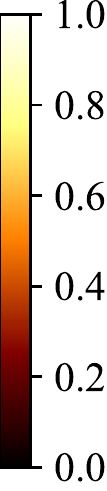}};
    
    \node[inner sep=0pt] () at (4.5, -10) {\includegraphics[width = 3.5cm]{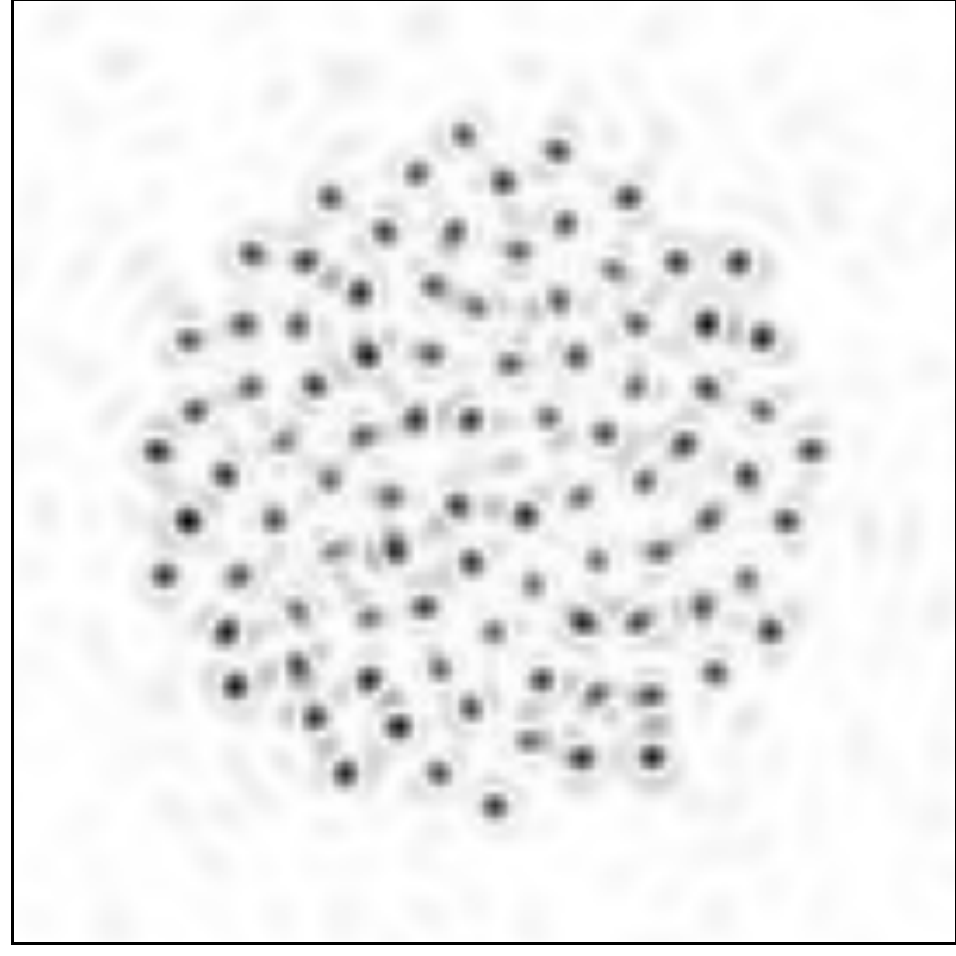}};
    \node[inner sep=0pt] () at (6.75, -9.9) {\includegraphics[height = 3.95cm]{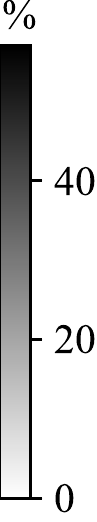}};
    
    \node[inner sep=0pt] () at (4.5, -13.85) {\includegraphics[width = 3.5cm]{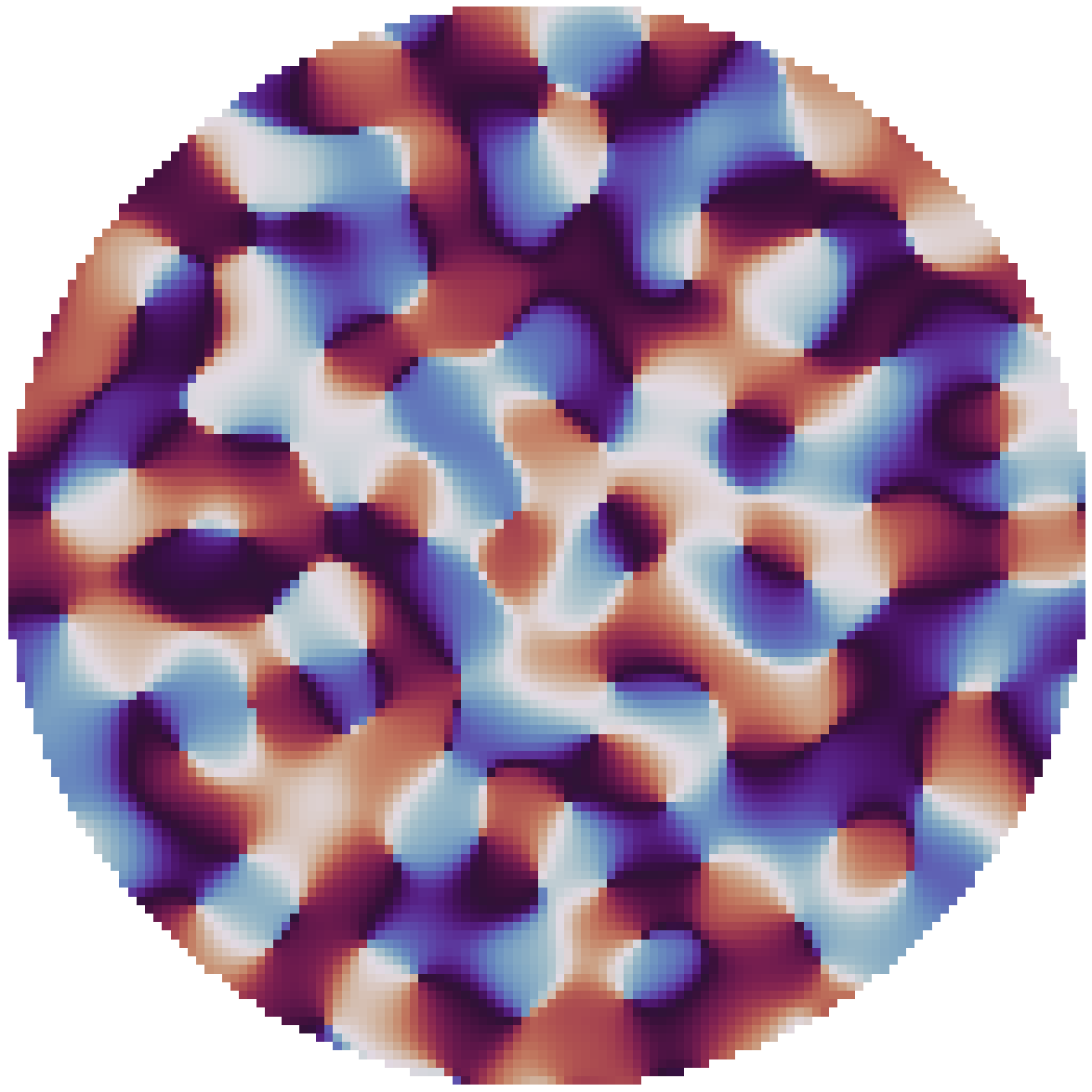}};
    \node[inner sep=0pt] () at (6.78, -13.85) {\includegraphics[height = 3.5cm]{figures/gradient_desc/toliman_true_pupil_cb.pdf}};
    
    \node[inner sep=0pt] () at (9, -2) {\includegraphics[width = 3.5cm]{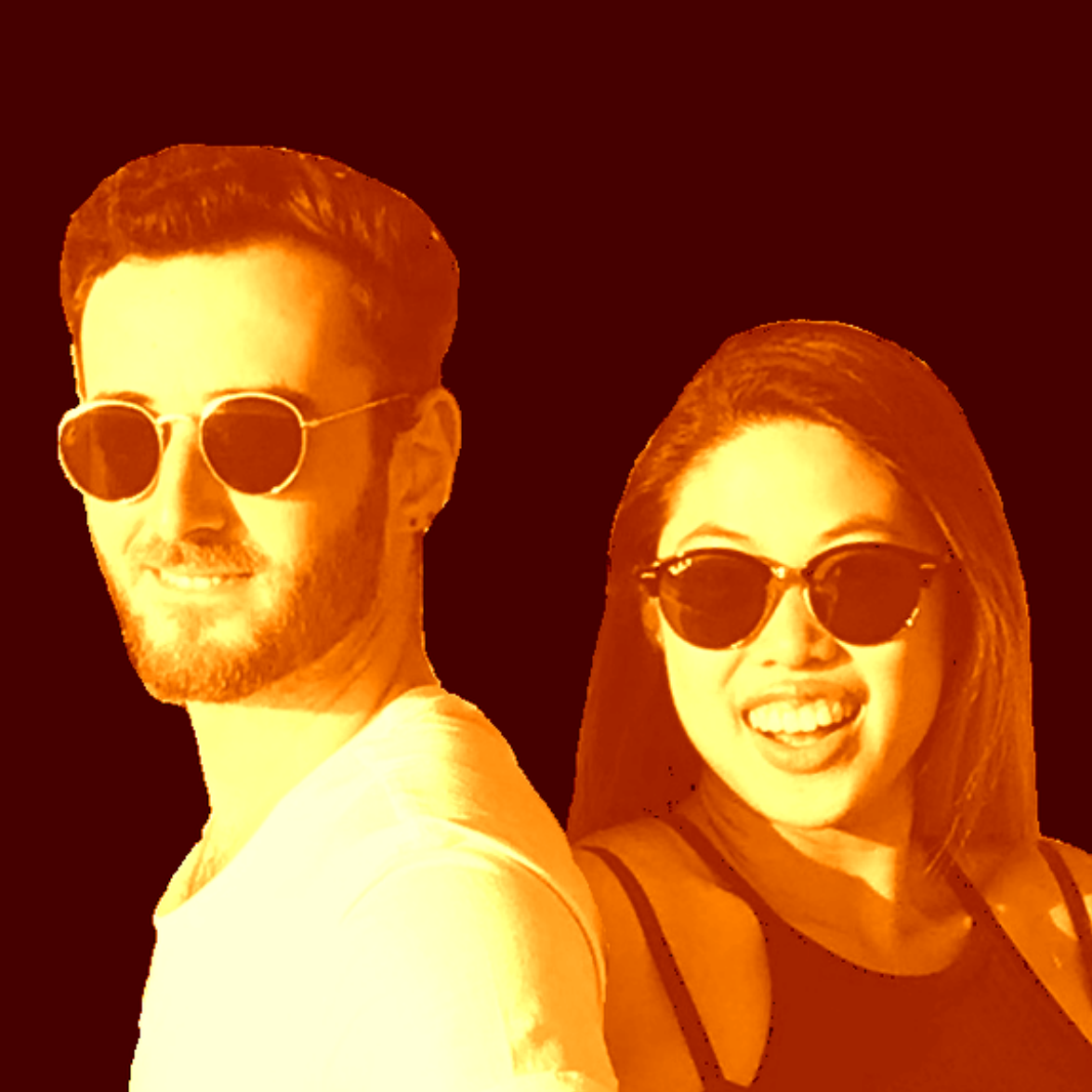}};
    \node[inner sep=0pt] () at (11.25, -2.05) {\includegraphics[height = 3.75cm]{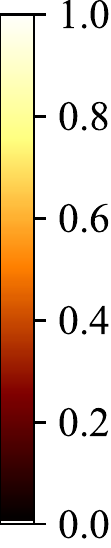}};
    
    \node[inner sep=0pt] () at (9, -6) {\includegraphics[width = 3.5cm]{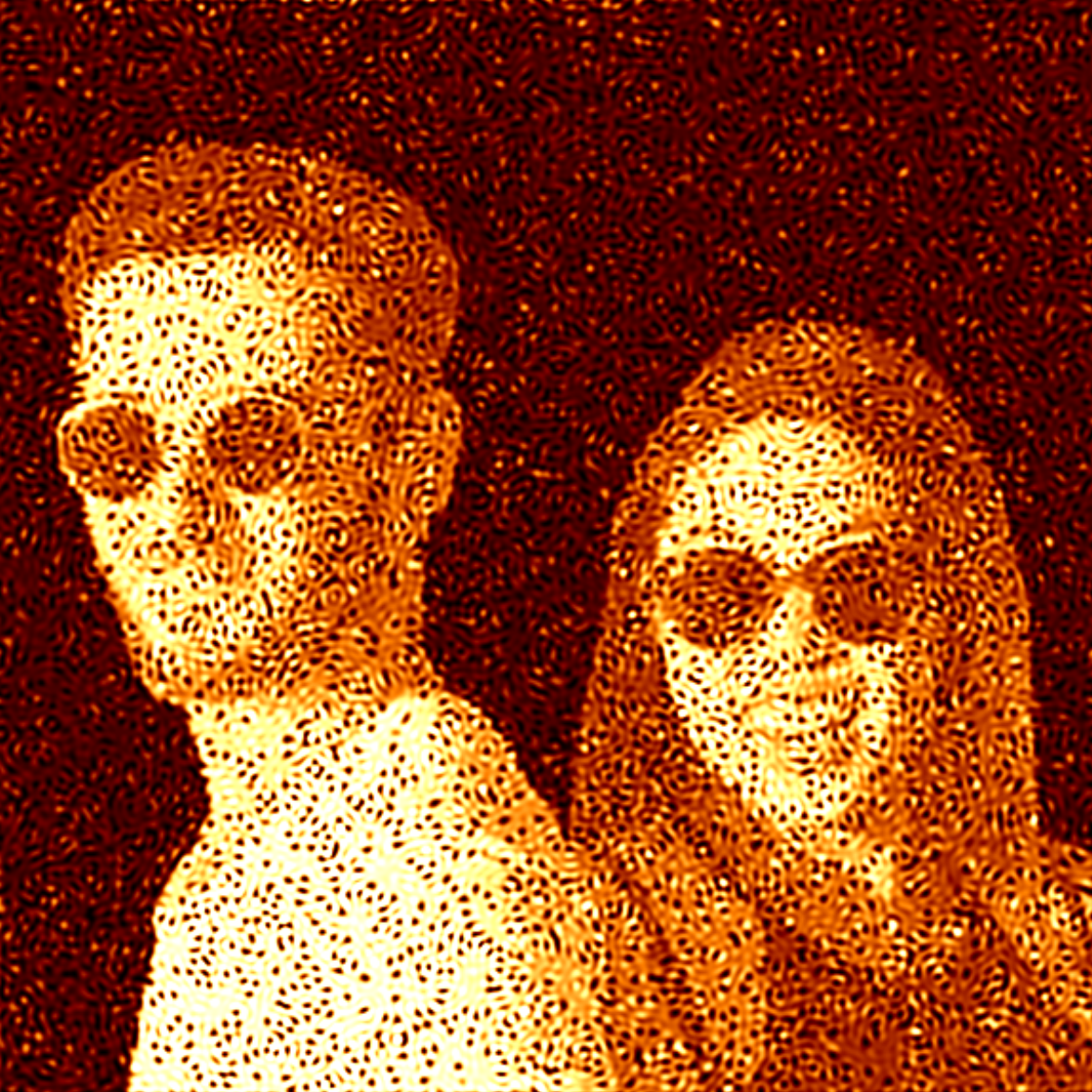}};
    \node[inner sep=0pt] () at (11.25, -6.05) {\includegraphics[height = 3.75cm]{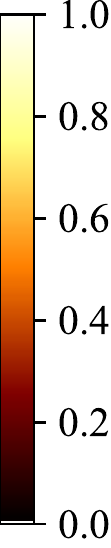}};
    
    \node[inner sep=0pt] () at (9, -10) {\includegraphics[width = 3.5cm]{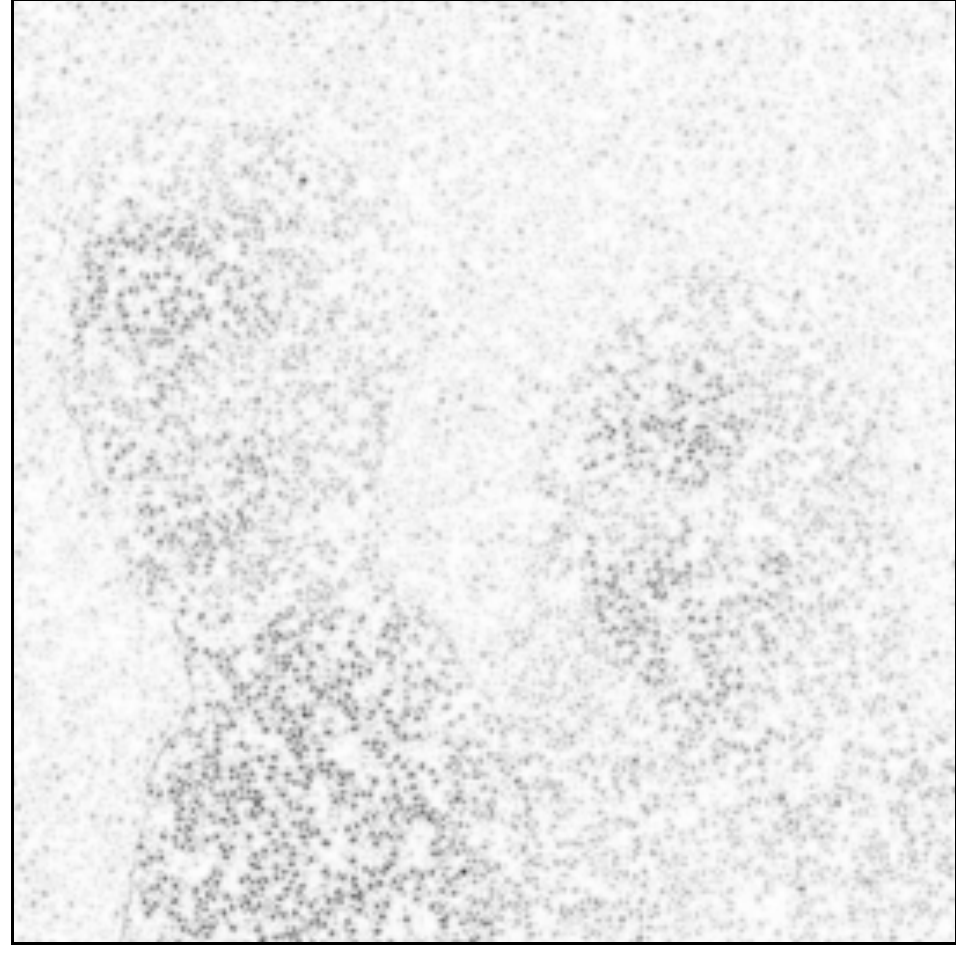}};
    \node[inner sep=0pt] () at (11.35, -9.9) {\includegraphics[height = 3.95cm]{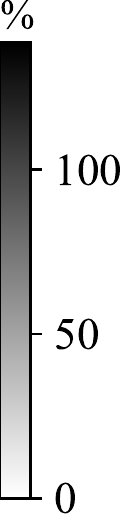}};
    
    \node[inner sep=0pt] () at (9, -13.85) {\includegraphics[width = 3.3cm]{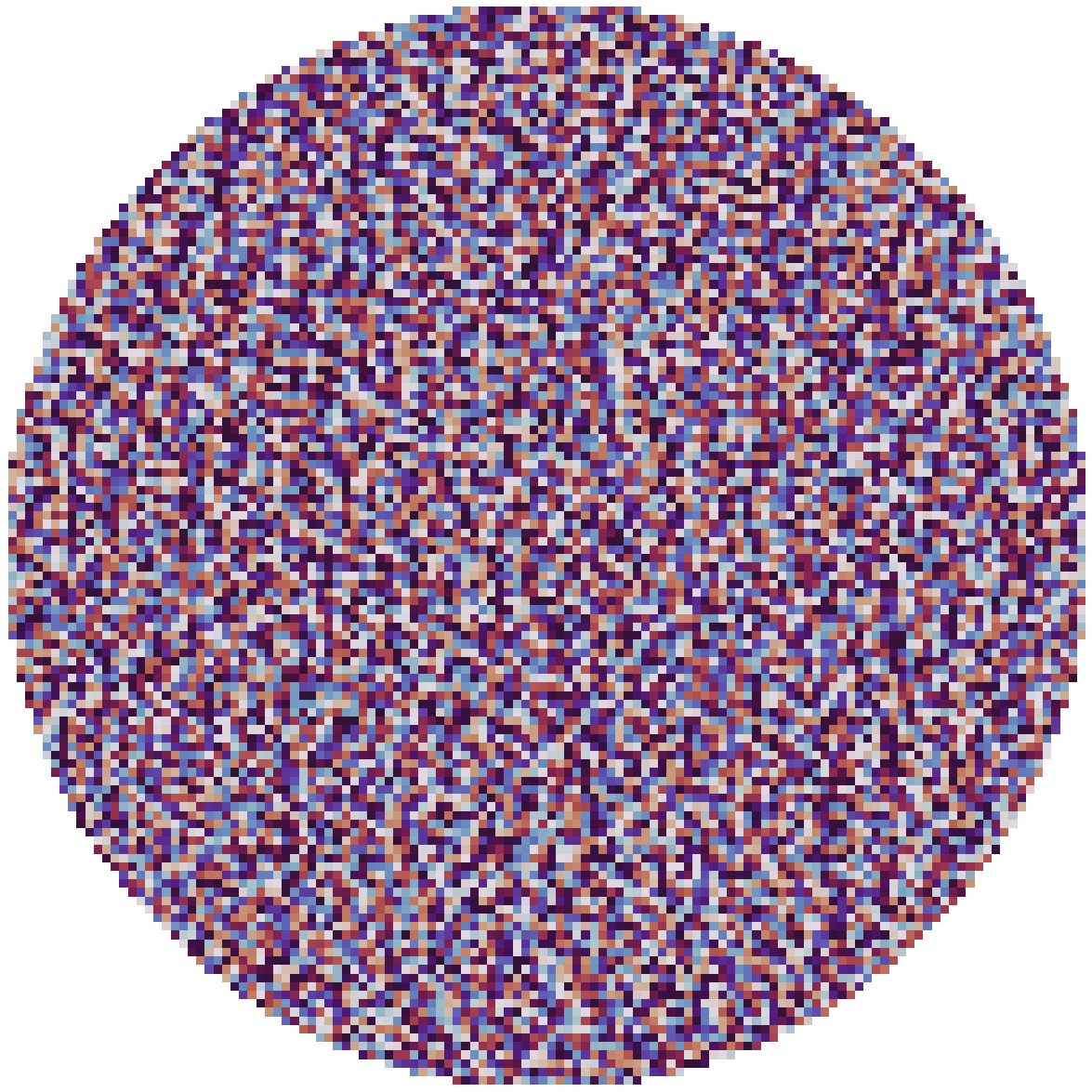}};
    \node[inner sep=0pt] () at (11.28, -13.85) {\includegraphics[height = 3.5cm]{figures/gradient_desc/toliman_true_pupil_cb.pdf}};

    \end{tikzpicture}
    
         \caption{Phase retrieval with gradient descent and the Adam optimizer for 3 cases. TinyTol: phase solution is known. Quantum Monodromy pattern: realistic PSF where the phase solution is unknown and may not exist. Faces of two co-authors: arbitrary PSF where the phase solution is unknown and may not exist. From top: target PSF, then reconstructed PSF, both normalized to maximum value; relative difference image; and recovered phase pupil. \href{https://github.com/alipwong/astronomical_phase_retrieval_and_design/tree/main/phase_retrieval/retrieval}{\color{linkcolor}\faGithub} 
         \label{fig:grad_desc_examples}}

\end{figure}

The first column in Figure~\ref{fig:grad_desc_examples} shows each of the target PSFs (assumed monochromatic). To recover the phase in the pupil plane, we initialized a pupil ($128 \times 128$ pixels) of small and random phase uniformly distributed $\in [-0.5, 0.5]$ radians from which the corresponding PSF ($256 \times 256$ pixels) was calculated and normalized to a total flux of 1. The objective was to minimize the error between the phase-estimated PSF and the target PSF. We used the \ac{mae} for the TinyTol and faces cases and the \ac{mse} for the QM pattern, which were experimentally found to work well. For the phase retrieval problem, \textsc{TensorFlow} was used to calculate the gradients of each pixel in the pupil plane with respect to the objective function, and the Adam \cite{kingma2014} algorithm was used to obtain an adaptive learning rate. While slower than the \textsc{Morphine} gradient descent implementation, \textsc{TensorFlow} was more convenient to later adapt for neural network implementations. The learning rates were initialized to $1$ (TinyTol), $10^7$ (QM pattern) and $10^3$ (faces), and were stepped down by a factor of 10 if the solution diverged. Additionally, the reconstruction of the QM pattern benefited from warm restarts where the learning rate was temporarily increased to drive the descent out of local optima. The recovered pupils are shown on the right, and the corresponding PSF shown in the center. 

Tasks solvable by gradient descent are also amenable to deep learning methods. Advantages of deep learning methods are that the neural networks can model more complex, nonlinear relationships and can optimize globally --- an improvement over gradient descent which is a local optimizer. Due to the connectivity of neurons in the network, the networks are able to optimize the pixels describing the phase in the pupil plane concurrently rather than independently. This approach of simultaneously optimizing structures at a range of scales has proven fruitful in mechanical design \cite{hoyer2019}, and we investigated whether this would bear similar advantages for multiscale optimization of phase masks. For comparison we implemented a dense neural network and a CNN, but found no benefit for these particular phase retrieval problems.

\subsubsection{Pupil recovery through Saturation}

Detector saturation is an issue that often needs to be considered in astronomical phase retrieval. This occurs when the flux count is too high for the detector and results in a nonlinear detector response --- typically observed at the central peak of the PSF. Autodiff phase retrieval with saturation has previously been demonstrated \cite{jurling2014}, and here we apply this to a severely saturated TinyTol PSF. The target PSF was the same as shown in Figure~\ref{fig:grad_desc_examples} but with the maximum value clipped to $5\times10^{5}$ counts in the unnormalized image. This corresponded to a maximum value of $1.2\times10^{-4}$ in the normalized (sum to 1) image and clipped the maximum value to $\sim23\%$ of the original maximum. A histogram of the pixel values in the PSF are shown in Figure~\ref{fig:saturation_example}. The demonstration shows how this method can accurately reconstruct pupil plane phases though nonlinear non-flux conservative processes such as pixel saturation.

Phase retrieval for the saturated PSF followed the same processes as outlined above but with modification to the objective function. Before calculating the \ac{mae} between the estimated PSF and the target PSF, the estimated PSF was clipped to the maximum value of the target PSF. Results are shown in Figure~\ref{fig:saturation_example}.

\subsubsection{Constrained Optimization}
Phase retrieval can benefit from domain knowledge which can be incorporated into the phase retrieval algorithm in the form of constraints. These can reduce the size of the search space making the optimization process more efficient, while also ensuring that the solution is realistic and physical. For example, this could be to meet fabrication requirements, impose symmetry or reflect prior knowledge of the optical system. In practice, any constraint can be used provided it is differentiable.

A practical example could be demonstrated in the reconstruction of the TinyTol pupil, which is known to be strictly binary. While it is not shown here, a constraint could be added to force phase values to $\pm \frac{\pi}{2}$. Additionally, symmetry constraints could be imposed, as TinyTol has 10-fold rotational symmetry.

For illustrative purposes, we demonstrate phase retrieval on the QM pattern where we constrain the phase $\phi$ to 4 evenly spaced values. The objective function used was 


\begin{equation}
		\cfrac{1}{N}\sum_{(\text{x, y})}\Big(\psi(\text{x, y}) - \hat{\psi}(\text{x, y})\Big)^2 + c \sum_{(p, q)} \Big| \Big(\phi(p, q) \ \text{mod} \ \frac{\pi}{2}\Big) - \frac{\pi}{4}\Big|.
	\label{eq:quadranise}
\end{equation}

\noindent Here the first term is the \ac{mse} of the target PSF $\psi$ and the estimated PSF $\hat{\psi}$, in dimensionless normalized counts and summed over pixels in the focal plane (x, y), and the second is the regularization term used to implement the phase constraint, where $(p, q)$ are the pixels in the pupil plane summed within the pupil support. The dimensionless regularization constant $c$ was used to adjust the relative weights of the two terms. This was initially set to $10^{-14}$ and increased every $\sim 1000$ epochs until it reached $2.5\times10^{-13}$. This allowed the algorithm to find an early fit to the PSF before slowly introducing the phase constraint.











\subsection{Discussion and Results}
\subsubsection{Pupil recovery by gradient descent}

Figure~\ref{fig:grad_desc_examples} shows successful recovery of the TinyTol pupil. This is perhaps unsurprising as the problem had a known solution. This solution was recovered in approximately $\sim$ 15 minutes on a MacBook Pro with 8 i9 cores and 32GB of RAM. To improve efficiency, symmetry and phase constraints could have been imposed. We attained a \ac{mae} in the PSF of $5.03\times10^{-11}$ and a \ac{mae} in the pupil of 0.0024 radians.

Although the TinyTol pupil contains only two possible phase states, we allowed phases to range continuously between 0 and $2\pi$ radians. Applying a constraint enforcing binarity on the recovered pupil results in a \ac{mae} in the pupil of $4.25\times10^{-8}$ radians. Given that the support consisted of 12874 pixels, a single flipped pixel (the minimum possible error) would contribute $2.4\times10^{-4}$ radians of error. This makes the error found in our binary-enforced pupil likely machine precision error and indicates that the recovery itself is essentially perfect.

Next we performed an investigation into phase recovery for a PSF where there is no known solution. We recovered the QM intensity pattern, which has speckled features characteristic of typical PSFs, although in this instance the replication of the target was imperfect.
Our algorithm's notable success was in yielding a PSF with the same structure as the QM pattern, however at some level of detail the peaks did not have as clean a structure as the target. This is seen in Figure~\ref{fig:grad_desc_examples} where the peak intensities did not match the target due to light bleeding between the peaks of the recovered PSF. This illustrates an intrinsic limitation: over a finite diameter pupil with an applied phase function, the space of solutions that can be obtained is bounded. The specific example here required generation of Gaussian function in the PSF; an outcome that cannot be obtained perfectly with a regular optical system. Despite this, the close approximation between the target and the recovered PSF points to practically valuable outcomes from this method.

\begin{figure}
    \centering
    \begin{tikzpicture}[scale = 1]
    
    \node at (-2.3, -2) [rotate = 90] {{Pupil}};
    \node at (-2.3, -6) [rotate = 90] {{PSF}};
    
    \node at (0, 0.2) [] {{Seed 0}};
    \node at (4, 0.2) [] {{Seed 1}};
    \node at (8, 0.2) [] {{Seed 2}};
    

    \node[inner sep=0pt] () at (0, -2) {\includegraphics[width = 3.5cm]{figures/gradient_desc/qm_pupil_nocb.pdf}};
    
    
    \node[inner sep=0pt] () at (4, -2) {\includegraphics[width = 3.5cm]{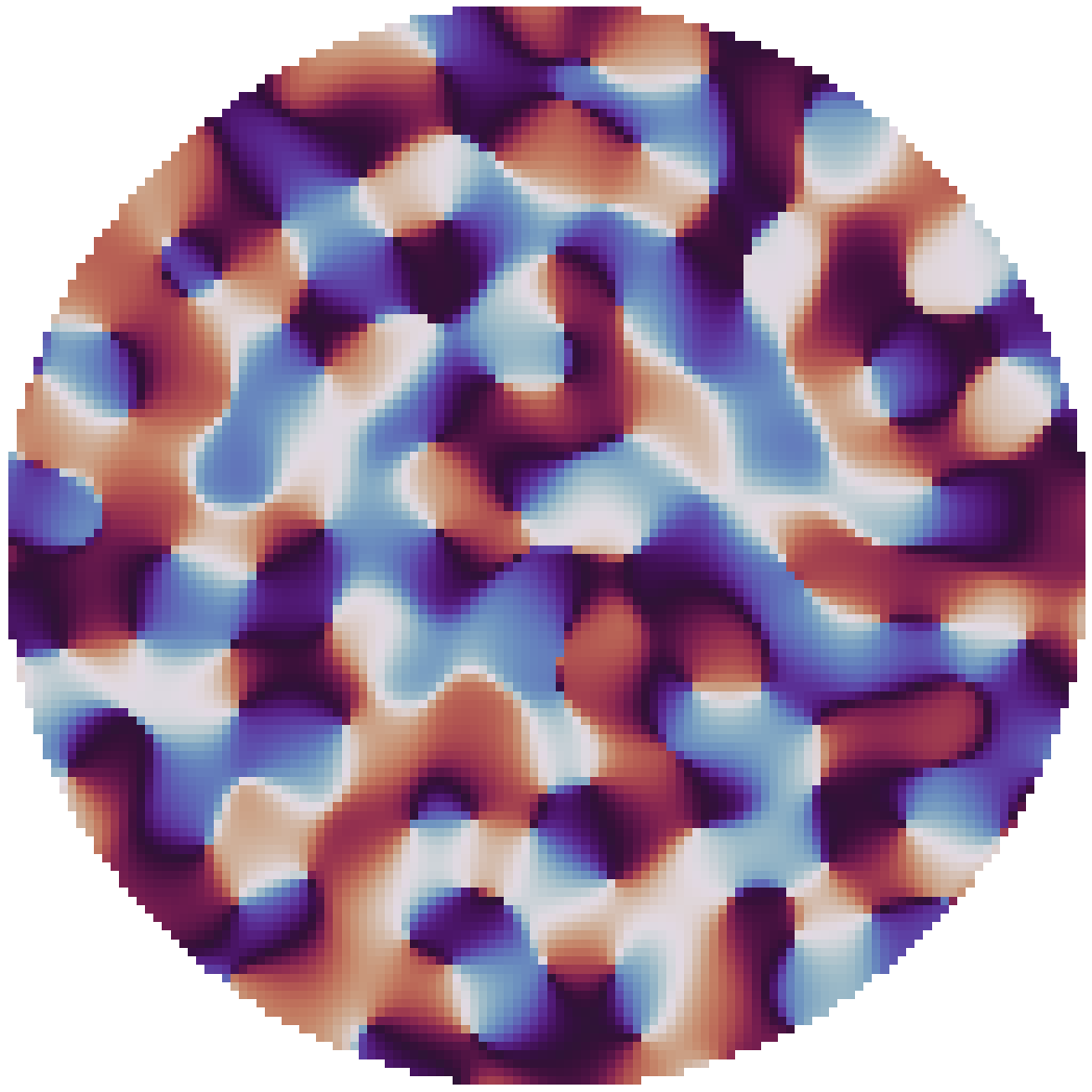}};
    
    
    \node[inner sep=0pt] () at (8, -2) {\includegraphics[width = 3.5cm]{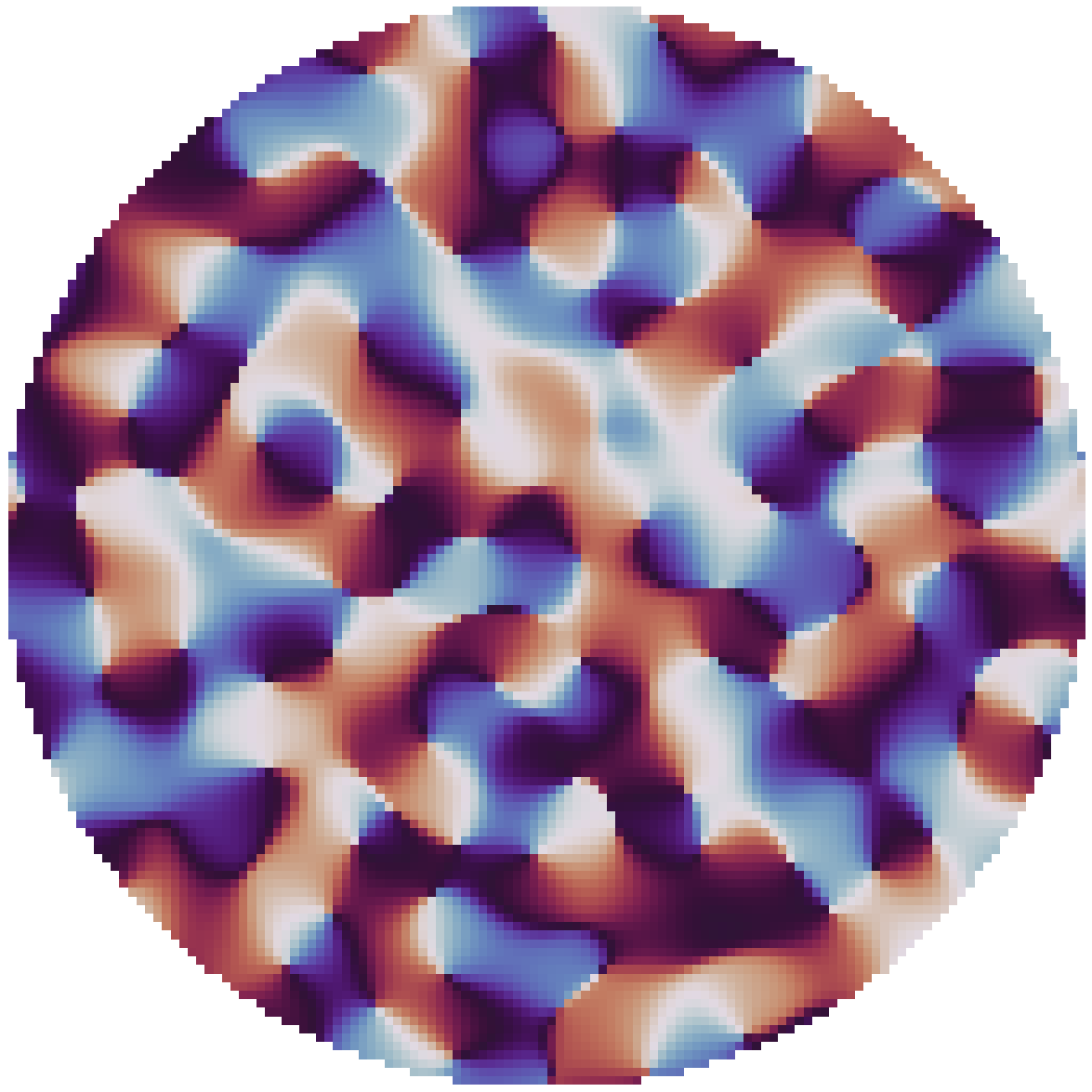}};
    
    \node[inner sep=0pt] () at (10.6, -2) {\includegraphics[height = 3.8cm]{figures/gradient_desc/toliman_true_pupil_cb.pdf}};

    \node[inner sep=0pt] () at (0, -6) {\includegraphics[width = 3.5cm]{figures/gradient_desc/qm_psf_zoom_nocb.pdf}};
    
    
    \node[inner sep=0pt] () at (4, -6) {\includegraphics[width = 3.5cm]{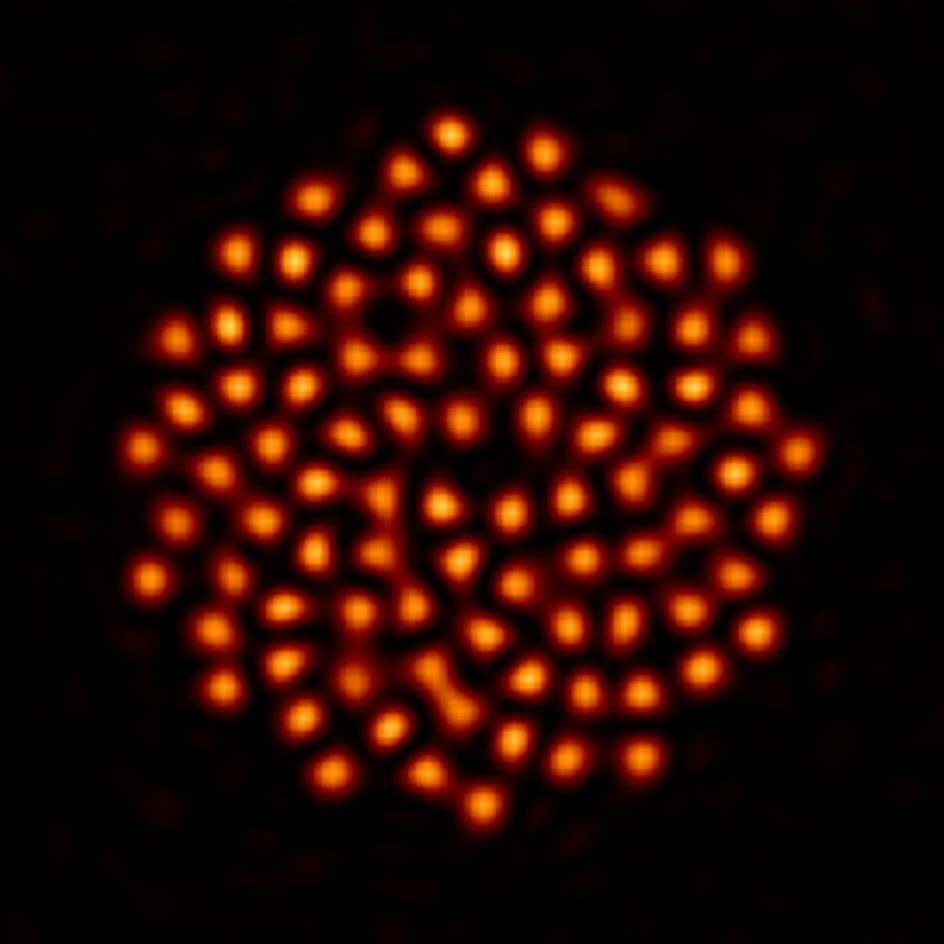}};
    
    
    \node[inner sep=0pt] () at (8, -6) {\includegraphics[width = 3.5cm]{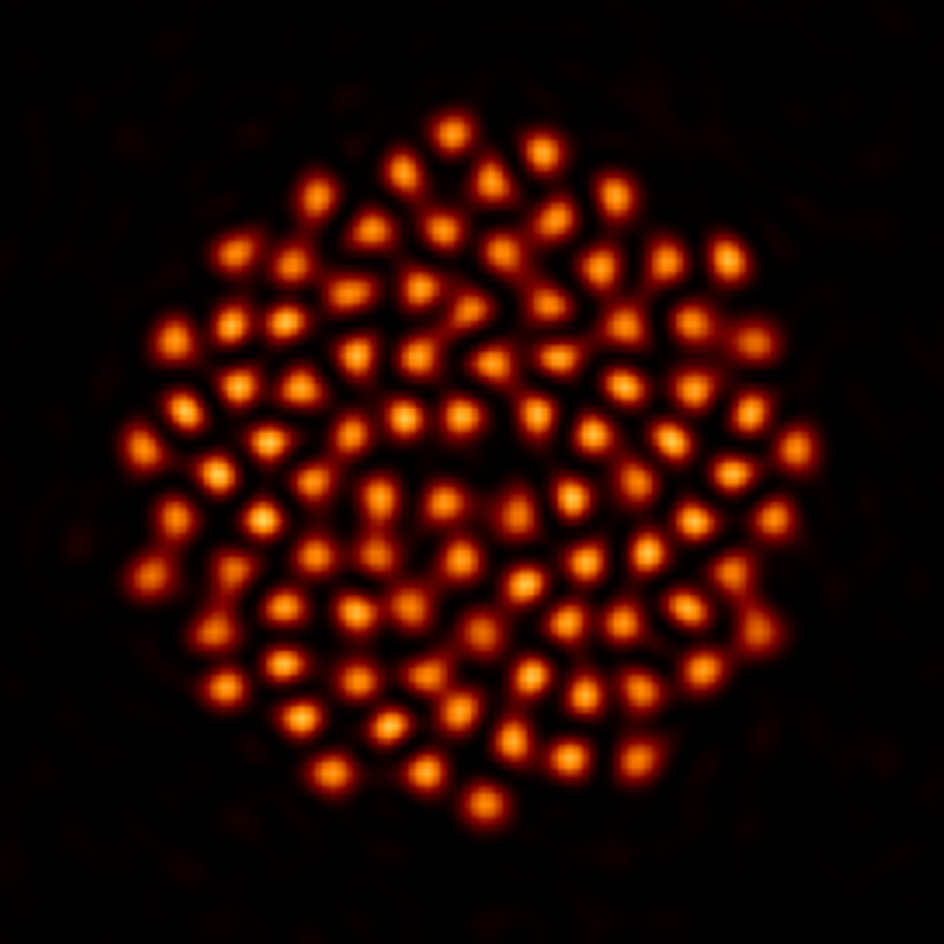}};
    
    \node[inner sep=0pt] () at (10.6, -6.05) {\includegraphics[height = 3.8cm]{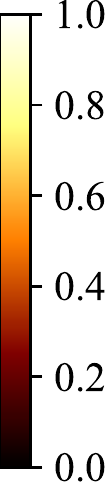}};

    \end{tikzpicture}
        \caption{Phase retrieval with gradient descent and the Adam optimizer on the Quantum Monodromy pattern with different initializations. Each presents an approximately equally good solution, with the mean absolute error (MAE) differing from case to case by a few parts per million. \href{https://github.com/alipwong/phase_retrieval_and_design/tree/main/phase_retrieval/seeds}{\color{linkcolor}\faGithub} \label{fig:qm_seeds}}

\end{figure}

Initializing the gradient descent with different seeds produced different, but equally good results (Figure~\ref{fig:qm_seeds}). This is indicative of a complex search space with numerous local minima, which explains why this descent benefited from warm restarts.

Our `faces' example presents a test case where the algorithm is presented an outcome that is demonstrably unobtainable. 
A perfect phase solution can not exist due to the spatial frequency structure of the target: both large smooth expanses and hard edges are problematic. The reconstructed PSF, although imperfect, is clearly recognizable to be human faces and key structural features of the original faces are clear. However, the overall texture is speckled, which is characteristic of the speckles observed in typical PSFs. As expected, the recovered pupil has high spatial frequencies in order to distribute light broadly across the PSF.


Additionally we attempted reconstructions with a dense neural network and a convolutional neural network for this same task, where the output of the networks were a proposed phase pattern, similar to methods used by Hoyer \cite{hoyer2019}. The results we obtained were marginally worse than those found via gradient descent and the models took longer to optimize. This was surprising as generally neural networks have the capacity to optimize globally and therefore should attain a better solutions. However, we believe that in practice, the added complexity of the networks hindered the optimization process and the increased complexity was unnecessary for the task.

While applying gradient descent to phase retrieval has been demonstrated previously, these results served as a baseline comparison for later experiments involving the handling of PSF saturation and the integration of domain knowledge in the form of constraints.

\subsubsection{Pupil recovery through Saturation}

\begin{figure}
    \centering
    \begin{tikzpicture}[scale = 1]
    
    \node at (-0.1, 5.15) [] {{True unsaturated PSF}};
    \node at (-0.1, 0.1) [] {{True saturated PSF}};
    \node at (4.6, 0.5) [] {{Reconstructed PSF}};
    \node at (4.6, 0.1) [] {{after clipping}};
    \node at (9, 0.1) [] {{Recovered pupil}};

    \node[inner sep=0pt] () at (-0.2, -2) {\includegraphics[width = 3.5cm]{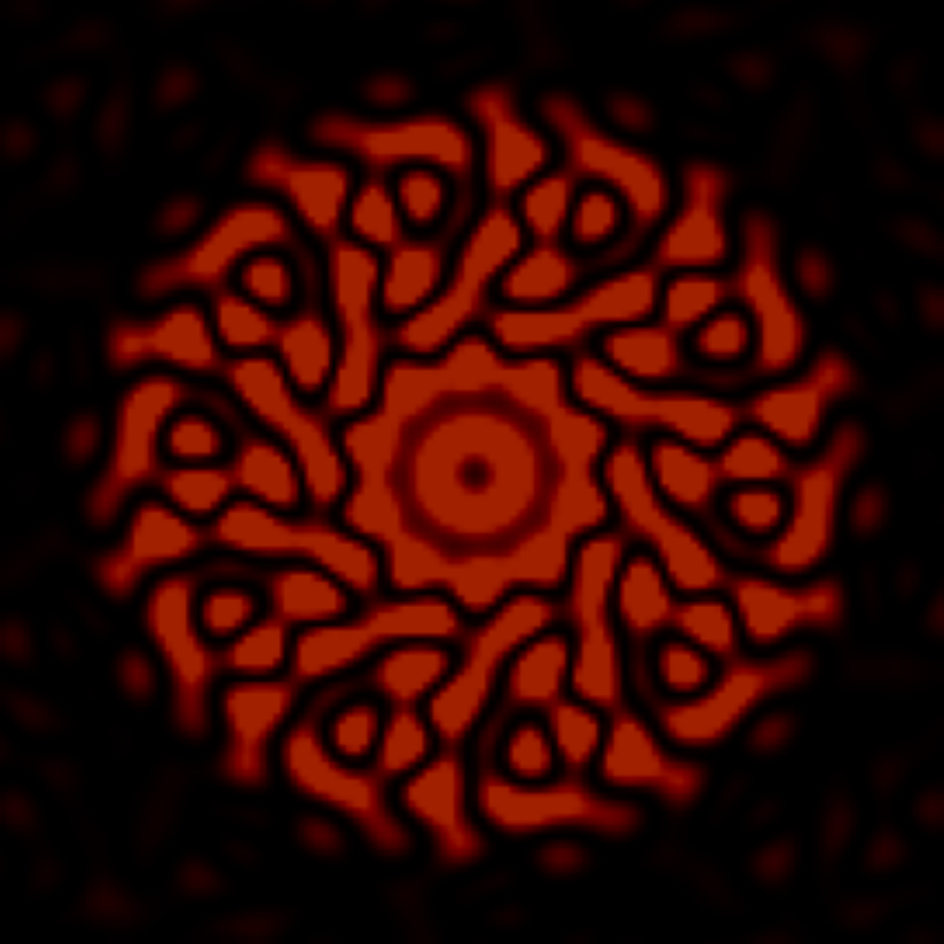}};
    
    \node[inner sep=0pt] () at (2.1, -2.05) {\includegraphics[height = 3.75cm]{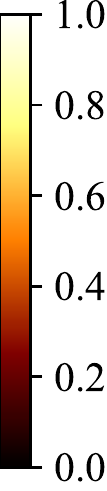}};
    
    \node[inner sep=0pt] () at (4.4, -2) {\includegraphics[width = 3.5cm]{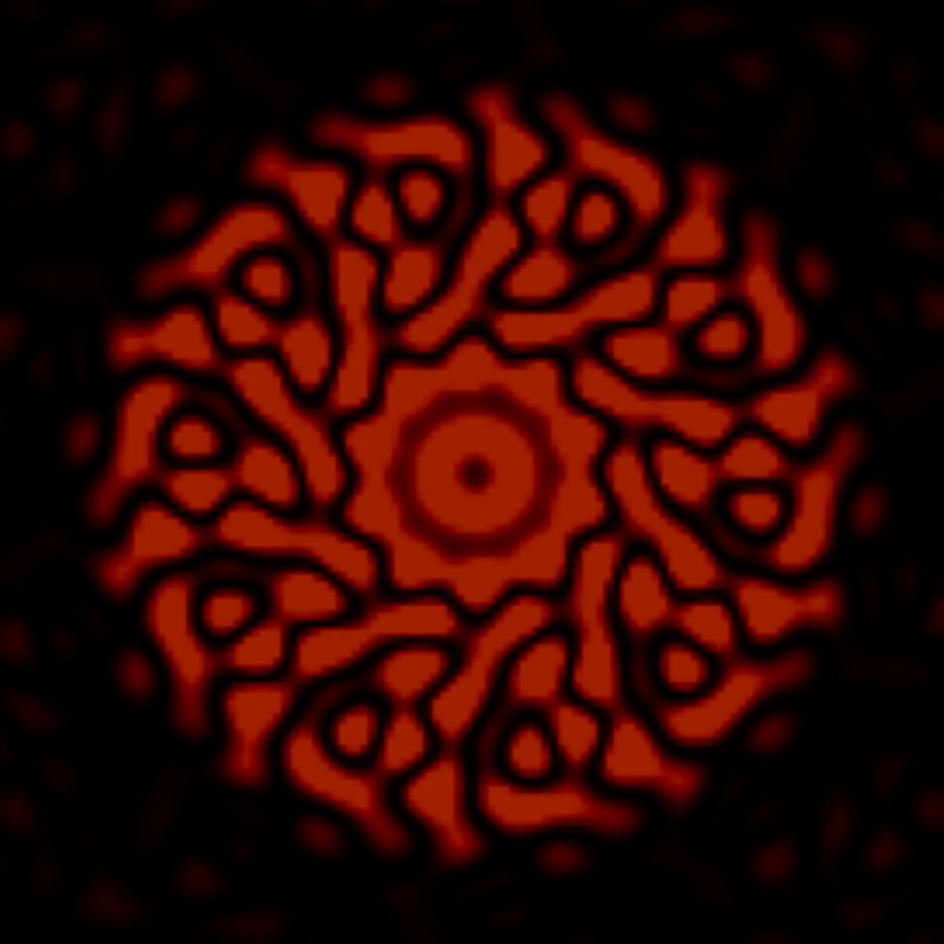}};
    
    \node[inner sep=0pt] () at (6.7, -2.05) {\includegraphics[height = 3.75cm]{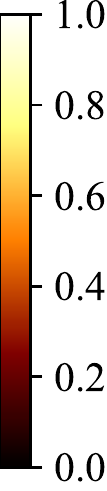}};
    
    
    \node[inner sep=0pt] () at (9.7, -2) {\includegraphics[height = 3.5cm]{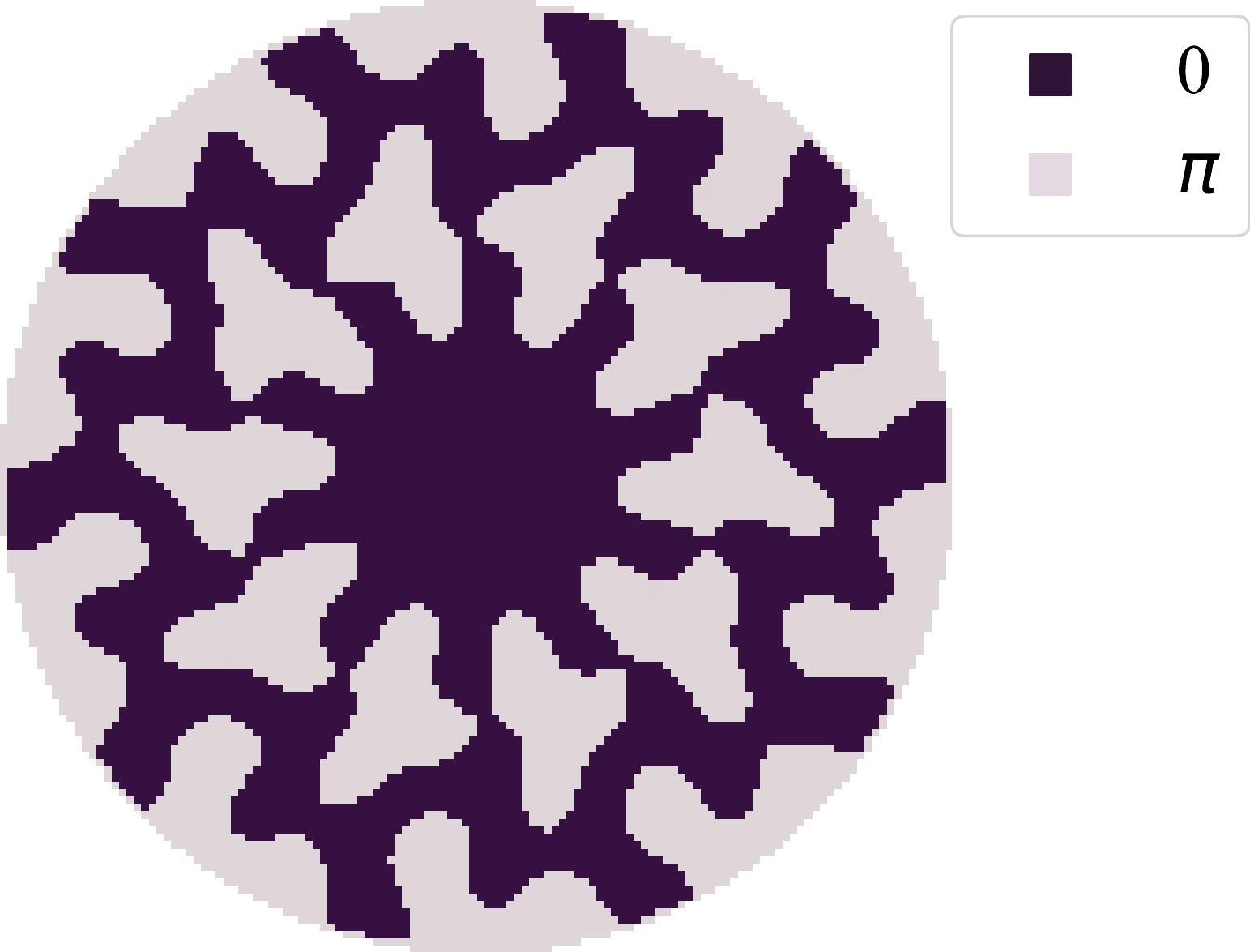}};
    
    
    \node[inner sep=0pt] () at (-0.2, 3.1) {\includegraphics[width = 3.5cm]{figures/gradient_desc/toliman_true_psf_zoom_nocb.pdf}};
    
    \node[inner sep=0pt] () at (2.1, 3.05) {\includegraphics[height = 3.75cm]{figures/gradient_desc/toliman_true_psf_zoom_cb.pdf}};
    
    \node[inner sep=0pt] () at (7, 2.75) {\includegraphics[height = 4.2cm]{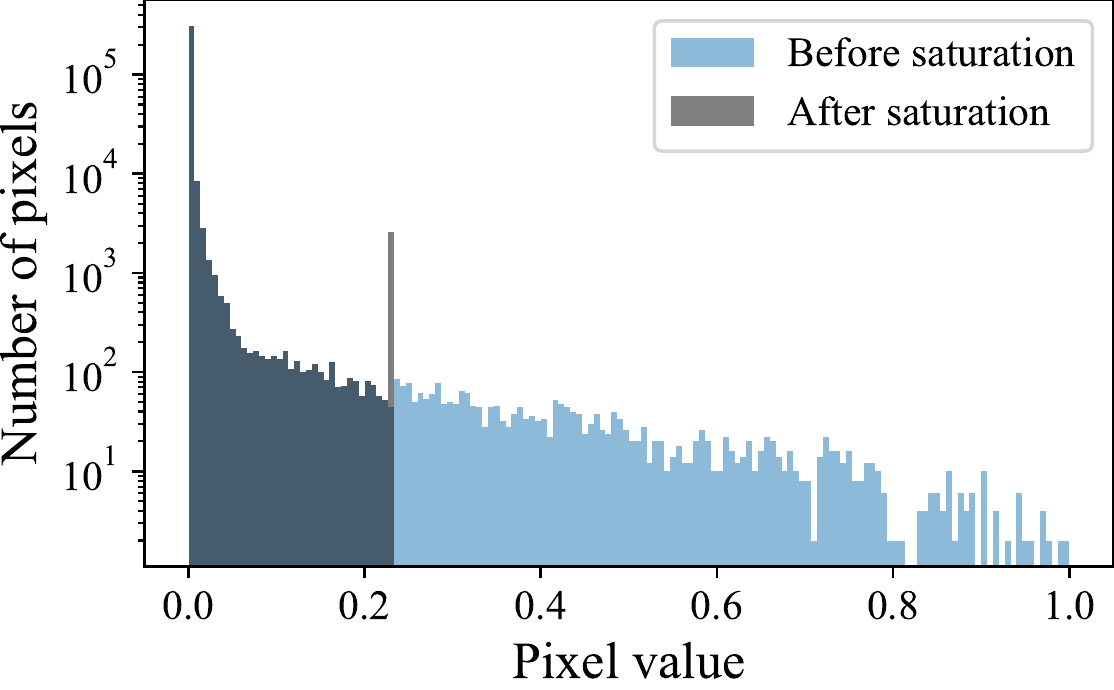}};
    
    \end{tikzpicture}
    
        \caption{\label{fig:saturation_example} Phase retrieval of a saturated PSF. The pixels in the true TinyTol PSF were saturated by clipping the values such that the maximum value was $\sim23\%$ of the original maximum. Phase retrieval was performed by comparing a clipped reconstructed PSF with the saturated PSF. \href{https://github.com/alipwong/phase_retrieval_and_design/blob/main/phase_retrieval/saturation/saturated.ipynb}{\color{linkcolor}\faGithub}}

\end{figure}

Figure~\ref{fig:saturation_example} shows the results of phase retrieval from a saturated PSF. We have found that we were able to successfully recover the TinyTol pupil achieving a \ac{mae} of $3\times10^{-10}$ in the PSF and $0.005$ radians in the pupil. Admittedly the error in the pupil is 2 times worse than in the unsaturated pupil. However, by applying a binary constraint to the reconstructed pupil reduces the \ac{mae} to $4.25\times10^{-8}$, which is on the scale of machine precision as before.


\begin{figure}
    \centering
    \begin{tikzpicture}[scale = 1]
    
    \node at (0, 0.2) [] {{PSF}};
    \node at (4.5, 0.2) [] {{Pupil}};

    \node[inner sep=0pt] () at (0, -2) {\includegraphics[width = 3.5cm]{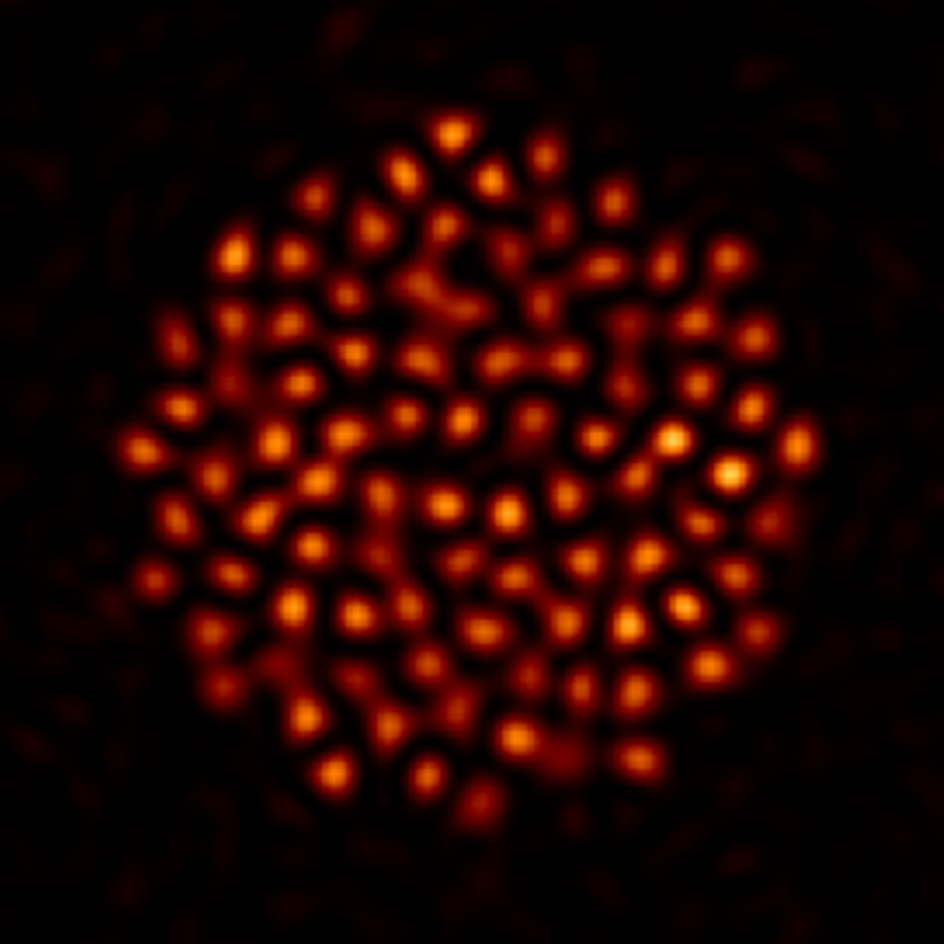}};
    \node[inner sep=0pt] () at (2.25, -2.02) {\includegraphics[height = 3.75cm]{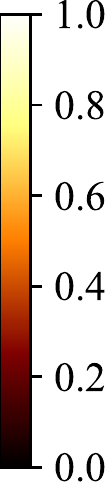}};
    
    \node[inner sep=0pt] () at (5.5, -2) {\includegraphics[height = 3.5cm]{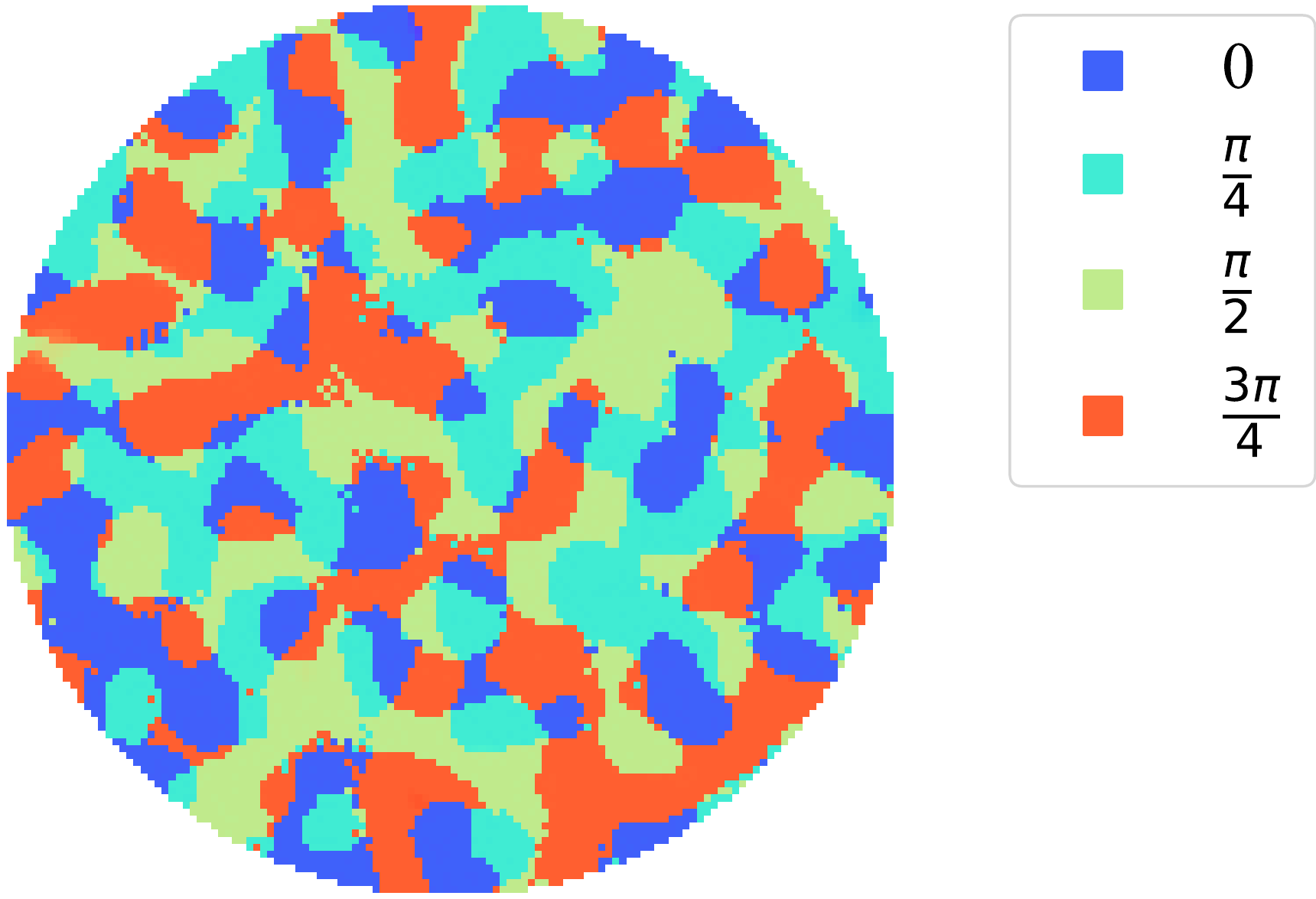}};
    
    
    \end{tikzpicture}
        \caption{Phase retrieval with the constraint given in Equation~\ref{eq:quadranise}. Here we enforce a 4-phase coloring on the pupil, which has less than a percentage increase in the \ac{mae} compared with the results in Figure.~\ref{fig:qm_seeds}. The constraint was not strictly enforced, but rather it was imposed by adding a penalty to the objective function for any deviations from the constraint. The constraint could later be enforced on the resultant pupil. For the example shown above, this would have negligible effect. \href{https://github.com/alipwong/phase_retrieval_and_design/blob/main/phase_retrieval/constraint/phase_constraint.ipynb}{\color{linkcolor}\faGithub}
        \label{fig:quadranise}
        }

\end{figure}

\subsubsection{Constraints}
Figure~\ref{fig:quadranise} shows the results of our reconstruction of the QM pattern when restricted to a 4-phase coloring. It is important to note that we did not enforce a strict phase constraint. The metric in Equation~\ref{eq:quadranise} simply penalizes phase that deviates from the 4 allowed phase values and the strength of the constraint is controlled by the coefficient $c$. If the final pupil does not result in a strict phase set, this can be enforced at a later stage by rounding the phase values. We perform this final step on our 4-phase QM pupil (Figure~\ref{fig:quadranise}) but it had negligible effects on the solution.

As expected, the 4-phase solution is not as good as the unconstrained solution shown in Figure~\ref{fig:qm_seeds} as the 4-phase constraint limits the complexity of the allowed pupil. The peaks are not as uniform and there is more bleeding, which is reflected in a slighty higher \ac{mae}, but this increase is less than a percentage.

In this demonstration, the phases were initialized with seed 0 and interestingly, the pupil maintains features evident in the seed 0 pupil shown in Figure~\ref{fig:qm_seeds}. This indicates that the algorithm search remains in the vicinity of the seed 0 solution, and never leaves. To escape local optima, the algorithm can be run with different initializations, or warm restarts applied during the fitting process. Given the similarity in quality of the solutions produced by different seeds, we believe that the problem is likely under constrained and multiple equally good solutions exist.


\bigskip
The examples shown in this section demonstrate how easily gradient descent can be applied to phase retrieval, while also illustrating the capabilities and limitations of this method in the presence of saturation or strong phase constraint conditions. It also lays down the necessary foundational work required to extend the phase retrieval method to the task of phase design.

\section{Phase Design}

We now consider an application of these gradient-descent methods to designing phase optics for astronomy, considering two examples: a scalar \ac{app} coronagraph (Section~\ref{sec:coronagraph}), and a diffractive pupil for astrometry (Section~\ref{sec:toliman}).

\subsection{Apodizing Phase-Plate Coronagraph}
\label{sec:coronagraph}
While \ac{app} coronagraphs are powerful tools for direct imaging, designing such systems is nontrivial. The difficulty of the problem is that the error space is non-convex, and iterative methods can struggle to escape local optima.  Previous approaches have applied phase iteration e.g. \cite{codona2004, codona2006}, modified Gerchberg-Saxton algorithms \cite{ruane2015}, or modified approaches to shaped-pupil coronagraph design modulating both phase and amplitude \cite{por2017}. These studies have shown that \ac{app} instruments are globally optimal for pupil-plane coronagraphs \cite{doelman2021}.

In this section we will show how gradient-based phase design provides a powerful and straightforward framework for optimizing a toy model: a monochromatic, scalar \ac{app} coronagraph with a multi-component objective function. We can apply gradient descent either to the coefficients in a modal basis, or to phase values on a pixelized grid, and straightforwardly achieve competitive coronagraph designs. 

We will use \textsc{Morphine} for optical simulation, which is a fork of the popular \textsc{Poppy} library \cite{poppy} using the \textsc{Jax} autodiff library \cite{jax} in place of \textsc{NumPy}. An advantage of this is that many instruments are already simulated in \textsc{Poppy}, which copes comfortably with multiple planes. These gradient descent methods are therefore easily extensible to designing hybrid coronagraphs, with phase and amplitude optics in conjugate planes.


\subsubsection{Constructing a Basis} \label{sssec:coronagraph_basis}

The simplest mode basis that can be used for phase design is the pixel basis, where each mode is the phase value of a single pixel in a grid. While this basis is complete up to high spatial frequency, it is quite inefficient for optimization as the dimensionality is $n^2$ for an $n \times n$ pixel pupil plane grid. A modal basis is often preferable as it does not scale with the grid size. 

It is often useful to construct a symmetric basis to match the symmetries of the pupil e.g. the rotational symmetry enforced by the spider arms that hold the secondary mirror. Further, symmetry reduces the complexity of the solution space and a lower-dimensional basis can be used to improve computational efficiency. Here we have designed a basis with 4-fold symmetry for a pupil support with 4 spiders.

\begin{figure}
    \centering
    \begin{tikzpicture}[scale = 1]
        \node at (0, -0.5) [] {Sample Logarithmic Radial Harmonic Functions (LRHFs)};
        \node at (0, -3.75) [] {Sample orthonormalized modes after performing PCA on the LRHFs};
        
    	\node[inner sep=0pt] () at (0, -2) {\includegraphics[width = 14cm]{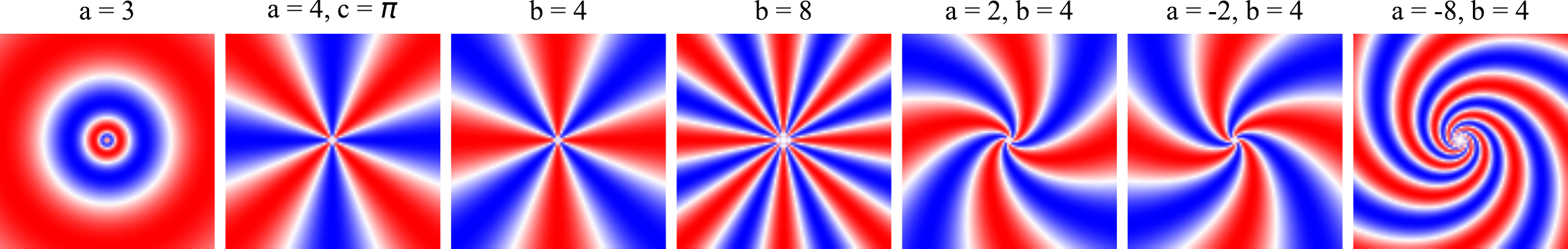}};
    
    	\node[inner sep=0pt] () at (7.5, -2.18) {\includegraphics[height = 2.1cm]{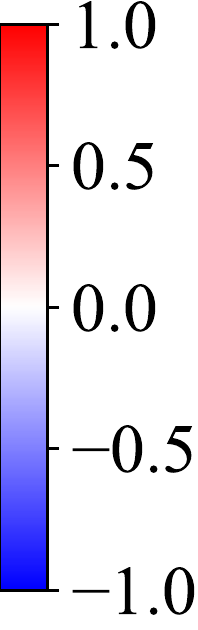}};
    	
    	\node[inner sep=0pt] () at (0, -5.25) {\includegraphics[width = 14cm]{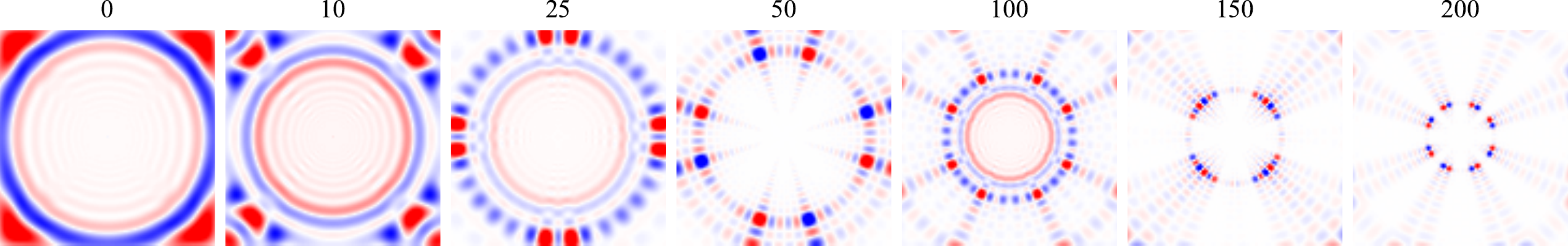}};
    
    	\node[inner sep=0pt] () at (7.5, -5.43) {\includegraphics[height = 1.95cm]{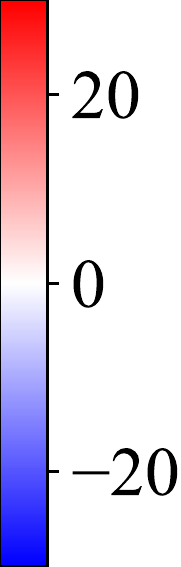}};
    	
    \end{tikzpicture}
    \caption{Top: Example modes created using Equation~\ref{eq:LRHF}. Constant values are 0 where not otherwise specified. $a$ controls the tightness of the spirals, $b$ controls the number of spokes and $c$ controls the rotation. Bottom: Sample modes after performing PCA. This was the mode basis used for coronagraph optimization.  \label{fig:LRHFs-mode-basis}}
   
\end{figure}

A basis was created using Logarithmic Radial Harmonic Functions (LRHFs) \cite{fletcher2003}, which were chosen as they contain a wide range of spatial frequencies and can therefore encode features at a range of scales and diffract light over a wide area of the focal plane. Each mode is defined such that:
\begin{equation}
	\phi(a, b, c) \equiv \cos(a\ln(r) + b\theta + c),
	\label{eq:LRHF}
\end{equation}

\noindent where $r$ is the radius, $\theta$ is the angle from the x-axis and $-30<a<30 \in \mathbb{Z}$, $b = 4B$ for $0<B<8 \in \mathbb{Z}$ and $0<c<2\pi$ are constants. Samples are shown in Figure~\ref{fig:LRHFs-mode-basis}. These modes have fine structure close to the origin that grows larger at greater radii. 

We combined the LRHFs with circular modes defining rings at equally spaced contours. These were defined by:
\begin{equation}
	\sin(r\pi i) 
	\label{eq:sin} 
\end{equation}
\noindent and
\begin{equation}
	\cos(r\pi i),
	\label{eq:cos}	
\end{equation}
\noindent for $0 < i < 50 \in \mathbb{Z}$.

The resultant basis was optimized by applying Principal Component Analysis (PCA) to reduce the basis down to 500 modes. Select modes from the optimized basis are shown in Figure~\ref{fig:LRHFs-mode-basis}.

The choice of basis was for demonstrative purposes only and is not necessarily optimal. In practice, the design of the basis should be informed by the task objectives.

\subsubsection{Algorithm}
Here we demonstrate na\"{i}ve \ac{app} coronagraph design with gradient descent for an annular coronagraph where the aim was to suppress light between 3 and 7 $\lambda/D$. We used the 500 mode basis shown in Figure~\ref{fig:LRHFs-mode-basis} and simulated an 8m telescope at a wavelength of 1.6 $\mu m$ with a detector pixelscale of $5\times10^{-3}$ arcseconds. The pupil plane is sampled on a $768\times 768$ pixel grid and the focal plane image $256 \times 256$ pixels. A 2.36\,m~radius secondary mirror obscures $\sim8.7\%$ of the pupil. The 4 spiders are placed at right angles to each other to maintain 4-fold symmetry.

The objective function balances two competing demands. The first is to minimize the peak value in the annulus and the second is to maximize the central peak. The objective function is then:
\begin{equation}
	a \times \max (\text{value in annulus}) - b \times \text{central peak},
	\label{eq:coronagraph} 	
\end{equation}
\noindent where $a$ and $b$ were constants used to weight the two metrics. Initially $b$ was large to enforce a strong central peak, otherwise the algorithm could resort to throwing light off the detector. $a$ and $b$ were tuned by hand during optimization, with the goal of maximizing $a$ without the solution diverging. We also hand-tuned the learning rate $\eta$, starting at $\eta = 10^{-4}$ and decreasing by a factor of 10 every few thousand epochs until reaching $\eta = 10^{-6}$. The PSF was normalized by dividing by the peak value in the Airy disk obtained from the unobstructed pupil (i.e. no secondary mirror or spiders).

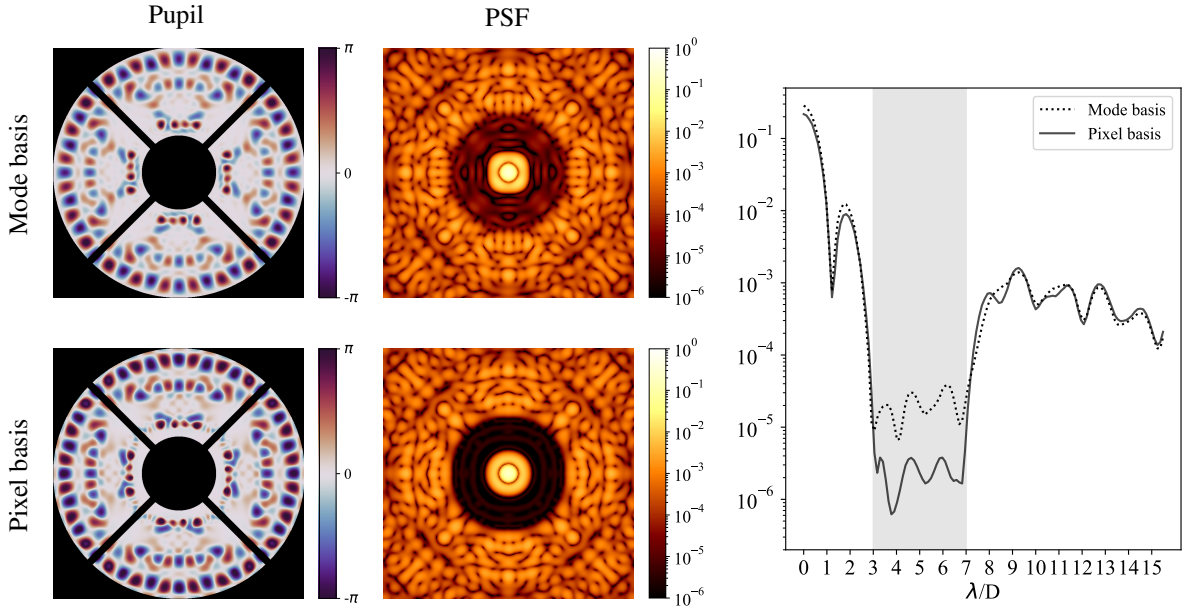
\begin{figure}
    \centering
    \begin{tikzpicture}[scale = 1]
    
        \node at (-0.4, 0.1) [] {{Pupil}};
        \node at (4, 0.1) [] {{PSF}};
        \node at (-2.5, -2) [rotate = 90] {{Mode basis}};
        \node at (-2.5, -6) [rotate = 90] {{Pixel basis}};

%
    	\node[inner sep=0pt] () at (0, -2) {\includegraphics[height = 3.5cm]{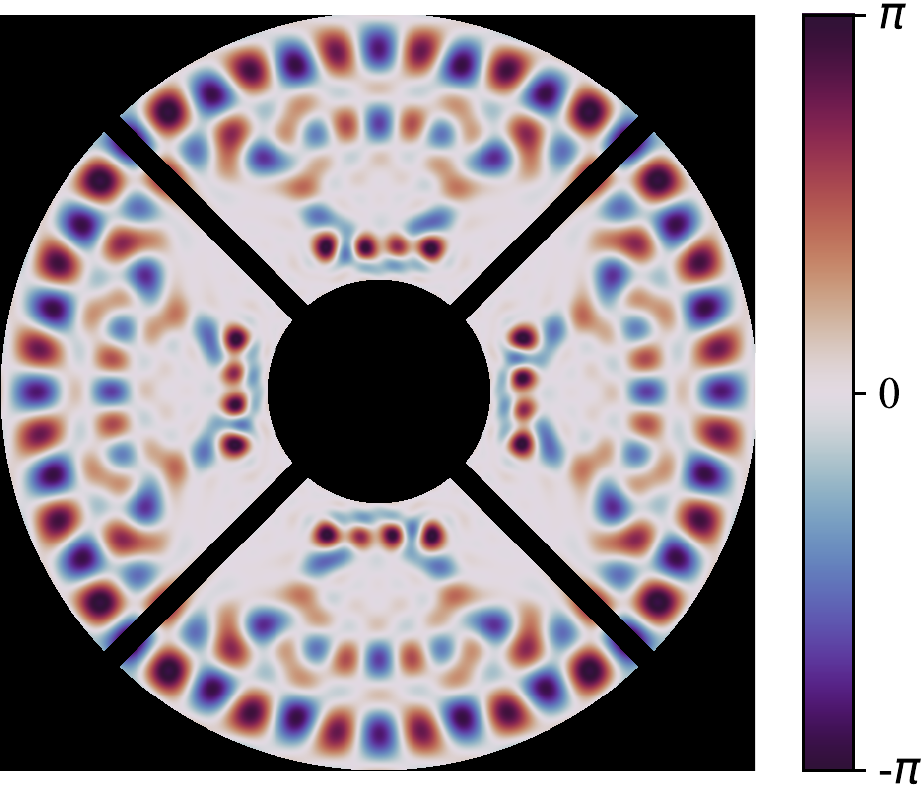}};
    
    	\node[inner sep=0pt] () at (4.5, -2) {\includegraphics[height = 3.7cm]{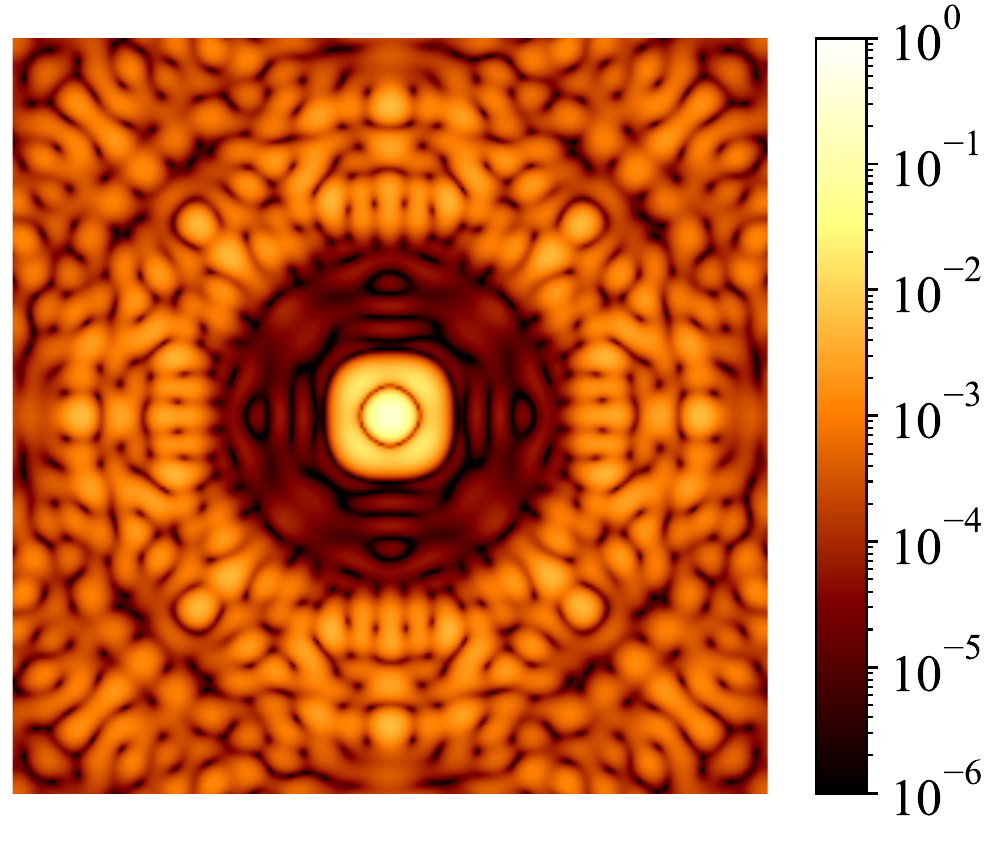}};
    	
    	
    	\node[inner sep=0pt] () at (0, -6) {\includegraphics[height = 3.5cm]{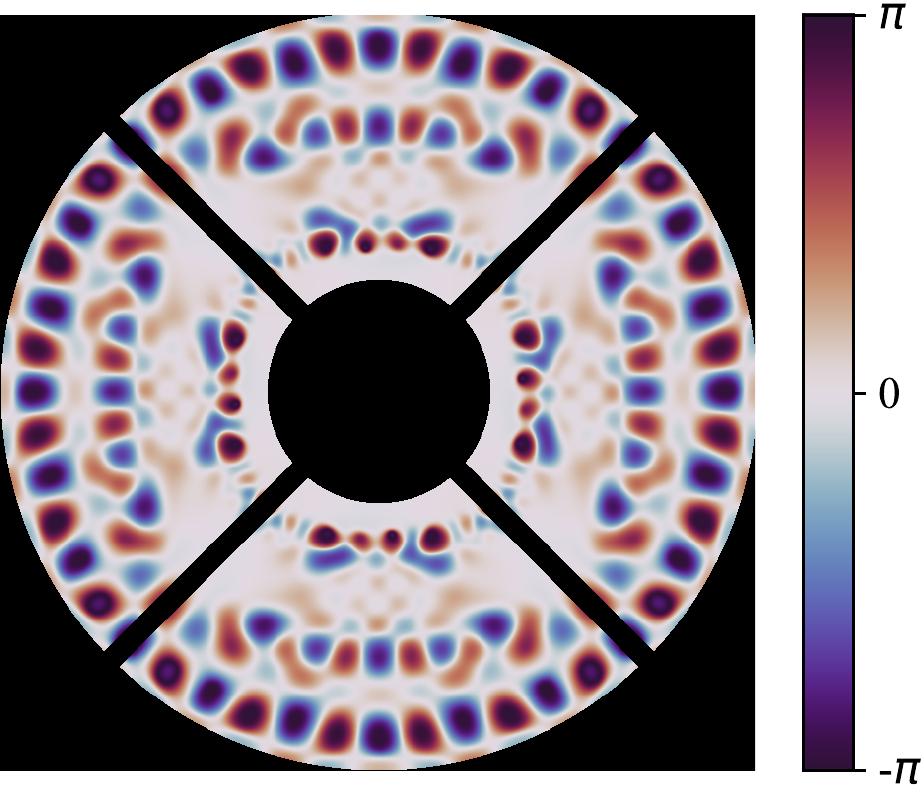}};
    
    	\node[inner sep=0pt] () at (4.5, -6) {\includegraphics[height = 3.7cm]{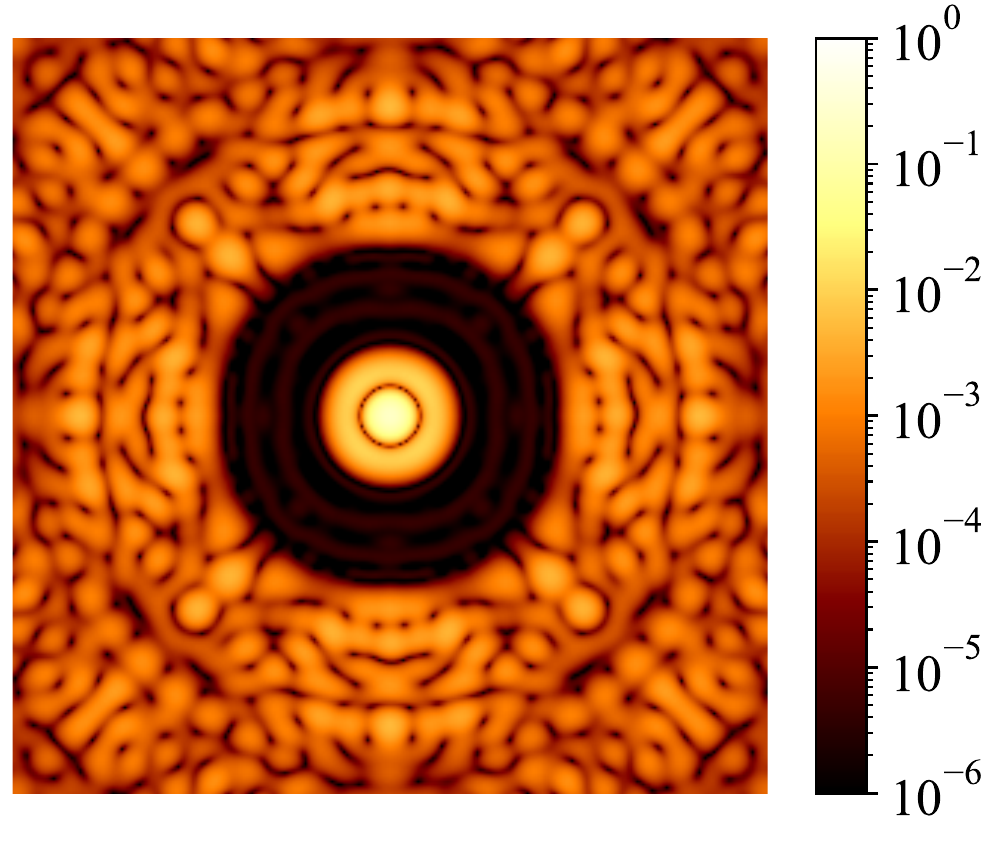}};
    	
    	\node[inner sep=0pt] () at (10, -4.28) {\includegraphics[height = 7cm]{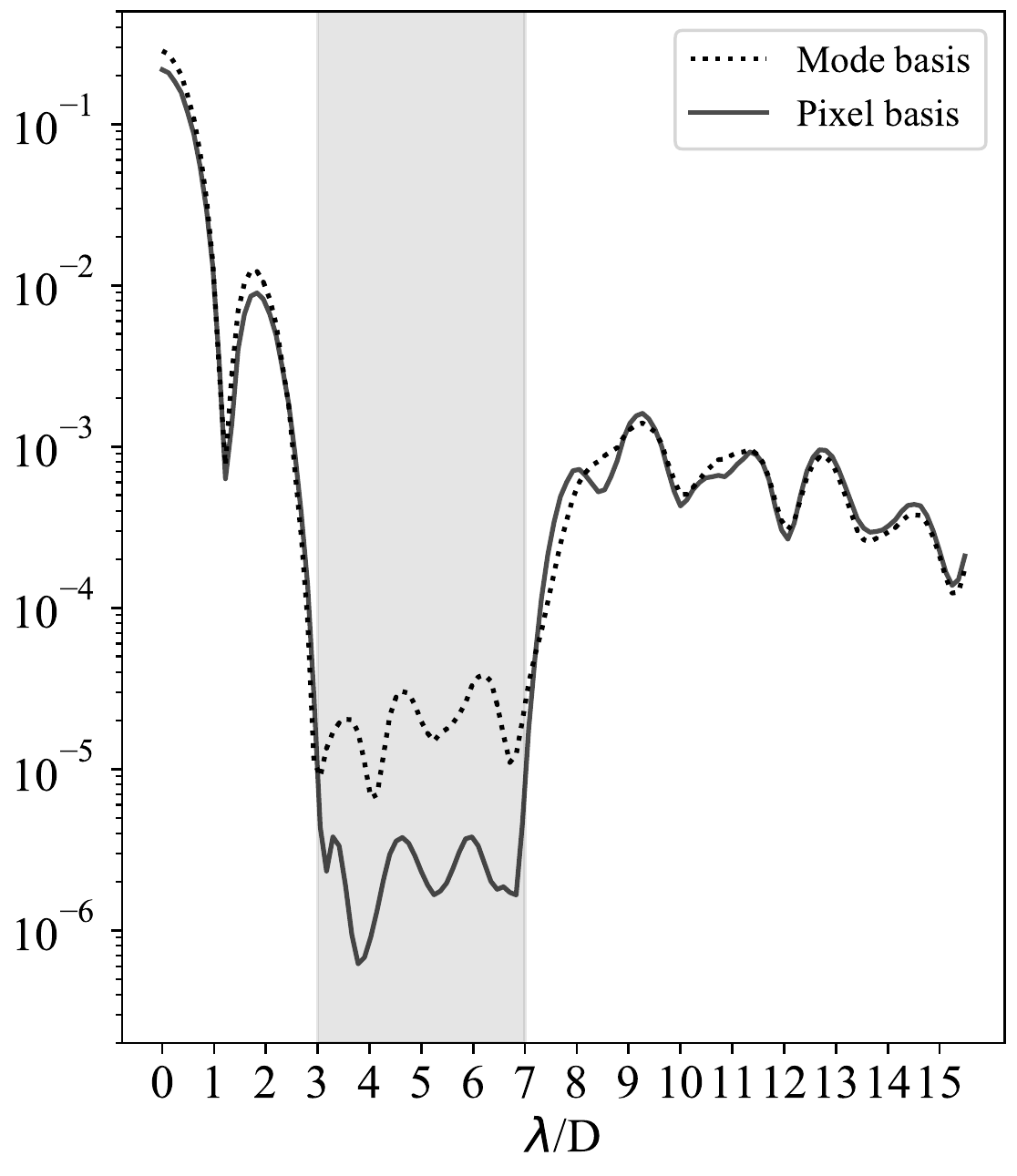}};
    	
    \end{tikzpicture}
    \caption{APP coronagraph optimization. Top left: coronagraph design and resultant PSF when optimizing using the modes shown in Figure.~\ref{fig:LRHFs-mode-basis}. Bottom left: Results of further optimization performed in the pixel basis initializing with the solution obtained in the mode basis. Right: comparison of the resultant PSFs. \href{https://github.com/alipwong/astronomical_phase_retrieval_and_design/tree/main/phase_design/coronagraph}{\color{linkcolor}\faGithub}\label{fig:coronagraph}}
\end{figure}

Optimization was first conducted in the mode basis as the 500 modes limited the complexity of the search space. The mode coefficients were initialized to small values uniformly randomly distributed~$\in [0, 10^{-3}]$ and gradient descent was performed on the mode coefficients. The results are shown in Figure~\ref{fig:coronagraph}.

Further optimization was conducted in the pixel basis. To maintain the 4-fold symmetry, only pixels in a single quadrant were optimized. For a pupil with $768\times 768$ pixels, a quadrant contained $384 \times 384 = 147~456$ pixels, although $\sim30\%$ were not in the pupil's support. The added complexity of the pixel basis allowed for a more optimal solution to be reached as shown in Figure~\ref{fig:coronagraph}. 

\subsubsection{Discussion and Results}

We found a two-stage approach to be effective: optimizing first in the symmetrized mode basis, before switching to the pixel basis. Using the mode basis we were able to efficiently navigate the simpler solution space and identify a promising solution. Transferring to the pixel basis then allowed the solution space to be more finely searched. Interestingly, converting to the pixel basis did not result in a large deviation from the general shape recovered in the modal basis.

During the fitting process, the learning rate and the metric weightings ($a$ and $b$ in Equation~\ref{eq:coronagraph}) were tuned by hand. $b$ was kept constant and $a$ was periodically increased to darken the region in the annulus. The purpose of the experiment was not to produce an optimal coronagraph, but to demonstrate that this method could be used for coronagraph design.

Table~\ref{tab:coronagraph} compares the coronagraph solutions from the mode basis and the pixel basis. While the pixel basis achieves a darker annulus, the mode basis does maintain a higher peak. We did not use a metric to identify the optimal trade-off between these terms in the objective function, but recognize that one could be used.

\begin{table}
\centering
\begin{tabular}{|c|c|c|c|}
\hline 
 & Peak & Mean in the annulus & 5$\sigma$ in the annulus\\ \hline
Mode basis & 0.284 & $2.677\times10^{-6}$ & $4.639\times10^{-5}$\\
Pixel basis & 0.217 & $2.975\times10^{-7}$ & $4.546\times10^{-6}$
\\ \hline	
\end{tabular}
\caption{Flux values are normalized to those simulated from an unobstructed pupil, i.e. no spiders or secondary. The Airy disk would have a peak value of 1.}
\label{tab:coronagraph}
\end{table}

\subsection{Diffractive Pupil}
\label{sec:toliman}

In this section we adopt the nomenclature for a pupil mask optimized for wide, astrometrically-useful PSFs to be a `diffractive pupil' \cite{guyon2012,tuthill2018}. These make an interesting test-case here due to the intrinsic complexity of the designs intended to produce relatively elaborate and strongly-featured PSFs. Here we apply our phase design methods to the diffractive pupil for the \toliman space mission. The original TinyTol pupil was designed using the Gerchberg-Saxton algorithm to try and achieve a patch of sharp peaks (a `bed of nails' pattern) by iteratively setting the amplitude of the complex wavefront to a top-hat function in the focal plane, and then re-applying the binary phase constraint in the pupil plane until convergence was reached. Here we performed optimization using the TinyTol constraints in order to compare the two optimization methods. We use 10-fold symmetry to match the existing TinyTol pupil.

One of the key differences between the TinyTol pupil and the \toliman pupil is that the \toliman pupil is designed for obstruction by a secondary mirror with radius 14.7\% of the primary and furthermore with spiders to hold this secondary mirror. However this mission is presently in design phase and the exact specifications are yet to be determined, and so a somewhat simplified system (for example omitting the spiders) was adopted here for illustrative purposes. We do however enforce 8-fold symmetry, which is appropriate for a system with 4 spider arms.


\subsubsection{Basis}

\begin{figure}
    \centering
        \begin{tikzpicture}

        \node at (0, 1) [] {\toliman};
        
    
    	\node[inner sep=0pt] () at (0, -0.7) {\includegraphics[width = 13cm]{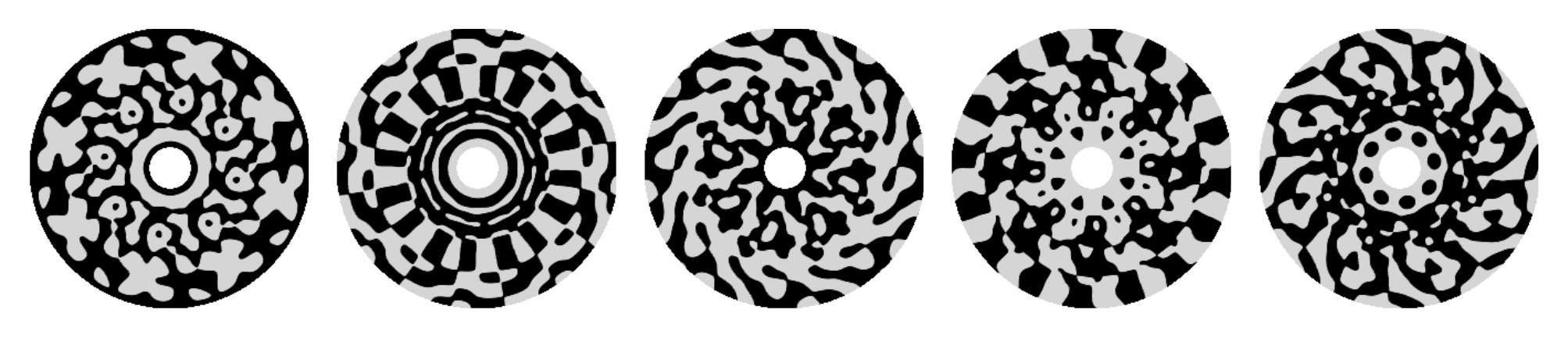}};
    	\node[inner sep=0pt] () at (7, 0.2) {\includegraphics[height = 1cm]{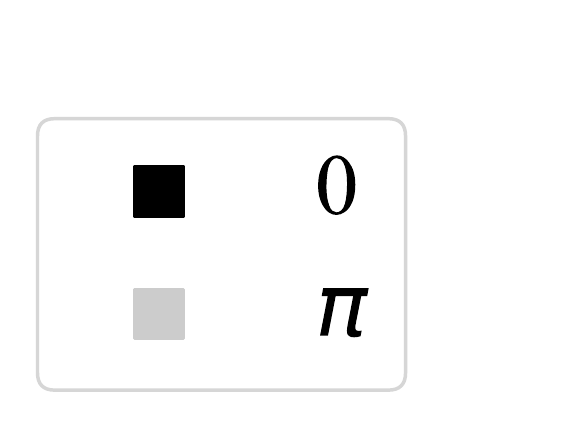}};
    	\end{tikzpicture}

    \caption{Randomly generated samples from the \toliman mode basis samples. The underlying modes are continuous, but to enforce binary phase, using the CLIMB algorithm positive values are mapped to $\pi$ and negative values are mapped to 0. The mode basis was convolved with a Gaussian of $\sigma = 7$ pixels before optimization to filter out the highest spatial frequencies.  \label{fig:tinytol_basis}}
   
\end{figure}

We constructed a basis with 10-fold symmetry to match the symmetry in the existing TinyTol pupil. This was constructed with the same methods as the basis used in coronagraph design outlined in Section~\ref{sssec:coronagraph_basis} but for 10-fold symmetry, again using PCA to optimize for the best 250 modes. Randomly generated samples are shown in Figure~\ref{fig:tinytol_basis} to illustrate the complexity of potential patterns spanned by the basis. In the design stage, the phases in the central region of the pupil (at $r < 0.35R_0$, where $R_0$ is the pupil radius) were set to 0 to avoid high frequency structures the develop toward the center of the pupil --- a consequence of the LRHFs.

The first phase in designing the basis for the \toliman pupil followed the same process but with 8-fold symmetry. However, in order to generate a pattern that could be more easily fabricated, additional steps were taken to remove some of the high frequency structures by convolving the mode basis with a Gaussian with $\sigma = 7$ pixels. This basis was then multiplied by the support before PCA was applied to retrieve the optimal 80 modes. A random sample of draws from this basis is shown in Figure~\ref{fig:tinytol_basis}.

\subsubsection{Methods: CLIMB}

In many cases it can be essential to design binary masks in phase or amplitude, whether for reasons of fabrication or physics. This is the case for the \toliman mission, and also for many kinds of the widely-used shaped-pupil coronagraphs, which are binary in amplitude \cite{tanaka2006}. Unfortunately derivatives can not be taken though discrete-valued functions as their gradients are necessarily either zero or infinity. So that we can design binary masks using the above symmetrized modal basis methods, we introduce an algorithm we call CLIMB: Continuous Latent-Image Mask Binarization. A latent phase pattern is generated as a sum of a continuous mode basis designed as above. The binary mask is then produced by thresholding this latent phase pattern to $\pi$ where it is positive, and 0 when negative. We can therefore continuously optimize over the coefficients in a small mode basis, while generating a high-resolution binarized mask.

In order to propagate derivatives through this thresholding, it is necessary to give the binary pattern soft edges, where the mask ansatz assumes a floating-point value $\in (0,1)$ only at the boundaries of regions. We describe this in pseudocode in Algorithm~\ref{climb_algo}. We therefore generate our continuous pattern on a $3N\times3N$ grid, which we downsample to the final $N\times N$ grid by applying the following prescription to each $3\times3$ subsampled pixel with phase values $z$. If all 9 pixels are positive, return $\pi$, or if negative, return zero. Otherwise we fit a 2D plane to the phase values, and calculate the line of intersection of this plane with the $z=0$ plane. We then return the fractional area of the unit square above this line, i.e. for which this locally linearized $\phi>0$. In practice, because this line can in general cut any two adjacent or opposite edges, this is first done by reorienting the square to a standard form and integrating. This results in a smooth transition from zero to unity as the edge of a binary region moves over a pixel, which means that gradients can propagate effectively even though the mask is zero or unity nearly everywhere.

\begin{algorithm}
\SetAlgoLined
\SetKwInput{Input}{Input} 
\SetKwInput{Output}{Output} 

\Input{$3N\times 3N$ pixel float-value map $Z$}
\Output{$N\times N$ soft-thresholded binary mask $\zeta$ which can propagate gradients}
\BlankLine
Initialize $\zeta$ empty.\\
\BlankLine
Fill each element $\zeta_{ij}$ by rebinning corresponding $3\times3$ block of $Z$:\\

\BlankLine
\For{\emph{each} $3\times3$ \emph{subgrid} $z \in Z$:}{
\uIf{\emph{all}(z)>0}{$\zeta_{ij} \leftarrow 1$}
\uElseIf{\emph{all}(z)<0}{$\zeta_{ij} \leftarrow 0$}
\Else{
\BlankLine
Fit a plane $ax + by + cz = 0$ by least squares:\\
\Indp
$a, b, c \leftarrow \text{LeastSquares}(z)$\\
\Indm
\BlankLine
Catch division by zero:\\
\Indp \lIf{\emph{any} $\{a,b,c\}_k = 0$}{
    $\{a,b,c\}_k \leftarrow \epsilon_\text{machine}$} 
\Indm
\BlankLine
Find intercepts of the nodal line with the unit square:\\
\Indp$x_1 = (-b-c)/(a)$\\
$x_2 = -c/(a)$\\
\Indm
\BlankLine
Reorient to a standard square:\\
\Indp $x_1, x_2 \leftarrow \min(x_1,x_2), \max(x_1,x_2)$\\
$x_1, x_2 \leftarrow \max(x_1,0), \min(x_2,1)$\\
\Indm
Calculate integral $A$ in standard form:\\
\Indp
$A \leftarrow x_1 - (c/b)x_2-(0.5 a/b)x_2^2  + (c/b) x_1+(0.5 a/b) x_1^2$\\
\Indm
\BlankLine
Catch edge cases involving inappropriate zero or NaN values.
\BlankLine
Enforce orientation:\\
\Indp
\eIf{\emph{mean}$({z}) >0$}{$\zeta_{ij}\leftarrow A$}{$\zeta_{ij}\leftarrow (1-A)$}
\Indm
}
}

 \caption{Continuous Latent Image Mask Binarization (CLIMB)}
 \label{climb_algo}

\end{algorithm}

\subsubsection{Diffractive Pupil Mask with CLIMB}

The goal was to design a diffractive pupil where the corresponding PSF was a circular flat top envelope around a `bed of nails' of sharp peaks, which are important for the \toliman diffractive-pupil astrometry concept. The key to this challenge was in the balancing of the different metrics.

To encourage peaks in the PSF, one objective function used was gradient energy (GE), given in Equation~\ref{eq:gradient_energy}. While it is important to maximize GE, by itself it is not a useful metric as maximum GE is achieved with an Airy disk. This is counter productive to the `circular flat top' requirement. Alternatively, a flat top radially weighted gradient energy (FTRWGE, Equation~\ref{eq:ftrwge}) can be used, which encourages peaks at larger radii up to the maximum radius $r_0$. However, used alone it too leads to global minima that are not ideal, tending towards ring-shaped PSFs of radius $r_0$. It was found that a combination of the GE and FTRWGE metrics performed well and resulted in a more even spread of light.
 

 \begin{equation}
 	GE = \sum_{(x, y)} \nabla \psi(x, y) = \sum_{(x, y)} \sqrt{ \Bigg(\frac{\partial \psi(x, y)}{\partial x}\Bigg)^2 +  \Bigg(\frac{\partial \psi (x, y)}{\partial y}\Bigg)^2}
 	\label{eq:gradient_energy}
 \end{equation}
 

 \begin{equation}
 	FTRWGE = \sum_{(x, y)} \sqrt{ \Bigg(\frac {\partial \psi(x, y)}{\partial x}\Bigg)^2 + \Bigg(\frac {\partial \psi (x, y)}{\partial y}\Bigg)^2} \cdot  \sqrt{x^2 + y^2} \cdot \rm{circ}(x, y)
 	\label{eq:ftrwge}
 \end{equation}
 
 \noindent where
 

\[ \rm{circ}(x, y) = 
    \begin{cases} 
      1 & \hspace{4mm} \text{if} \ \sqrt{x^2 + y^2} \leq r_0 \\
      0 & \hspace{4mm} \text{if} \ \sqrt{x^2 + y^2} > r_0 \\
   \end{cases}
\]
 

\noindent and $\psi$ is the PSF as defined earlier. In conjunction to the GE metrics, additional metrics were used to enforce the circular flat-top shape of the PSF. The radius of the main disk was approximately 10$\lambda$/D, in accordance with the existing TinyTol pupil.

\begin{figure}
    \centering
	\includegraphics[width = 6cm]{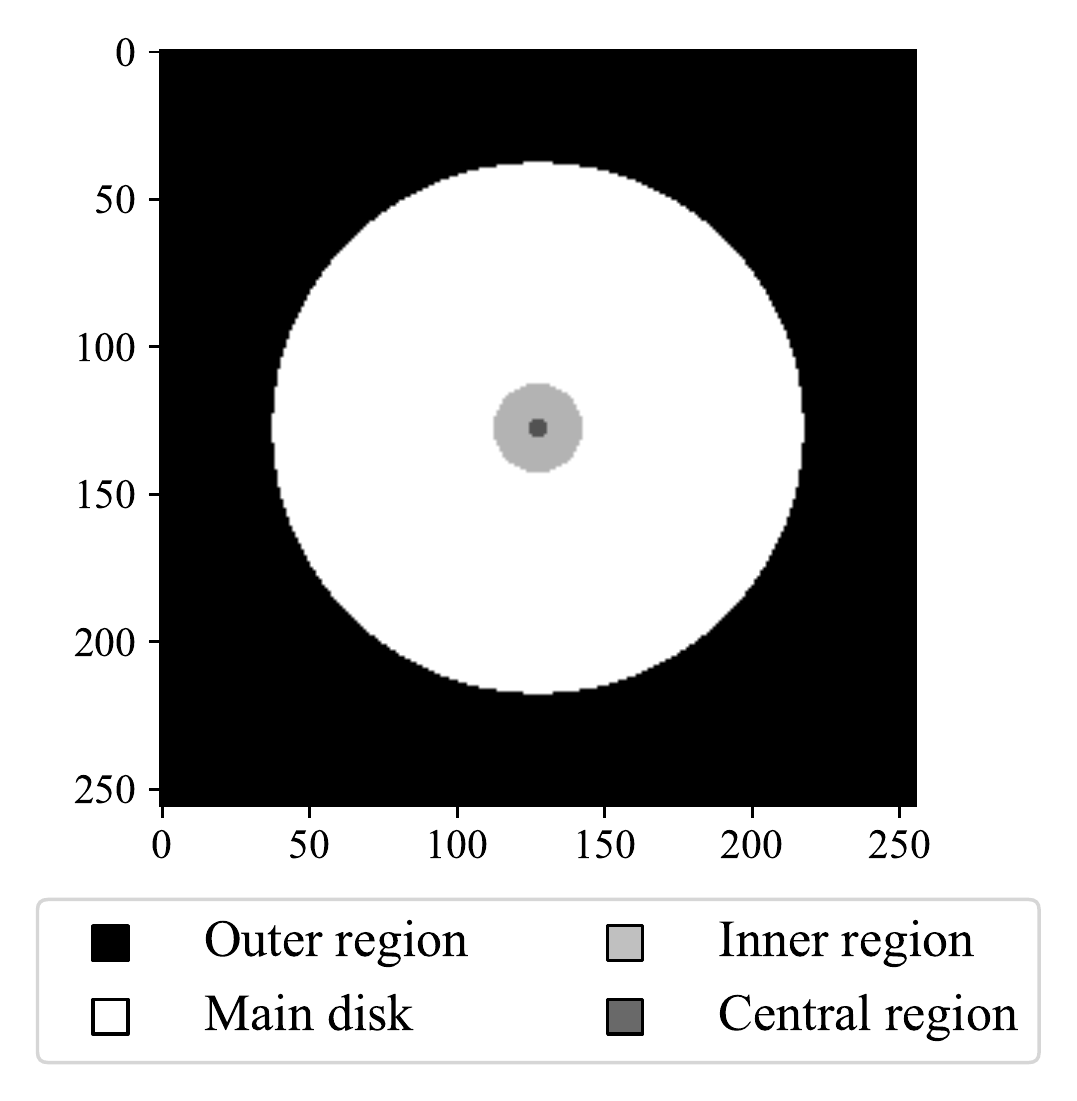}
    \caption{Defined regions referred to in Equations~\ref{eq:tinytol_metric} and~\ref{eq:toliman_metric}. Ideally the TinyTol and \toliman PSFs should have uniform peaks in the main disk and be suppressed in the outer, inner and central regions.}
    \label{fig:mask}
\end{figure}

A summary of metrics is given below, and were used where the PSFs were normalized to a total flux of 1. Figure~\ref{fig:mask} shows a map of the regions.

\begin{itemize} \itemsep 0em
	\item \textbf{Metric 1: Reduce light in the outer region}. This ensure maximum light in the central PSF and reduces broadband effects. $m_1$ was defined as the largest value in the outer region.
	\item \textbf{Metric 2: Reduce light in the inner region}. This ensures more light in the disk of the PSF and also helps to account for leakage, which will result in a central peak. $m_2$ was defined as the largest value at the center.
	\item \textbf{Metric 3: Reduce light in the central region}. This also contributed to ensuring an even spread of light around the disk. Otherwise solution tended toward enforcing the first Airy ring. $m_3$ was defined to be the largest value in the central region.
	\item \textbf{Metric 4: Reduce the maximum peak in the main disk}. This was to keep the light evenly spread and ensure that no subset of speckles dominated. $m_4$ was defined to be the largest value in the circular disk.
	\item \textbf{Metric 5: High gradient energy}. Combined the GE and FTRWGE equations.
\end{itemize}

Ultimately the final objective function used for optimizing the 10-fold symmetric TinyTol pupil was:
\begin{align}
	\mathscr{L}_\text{TinyTol} &= 15 \max(m_1, 0.0573) \nonumber \\
	& + 5\times10^{3} \max(m_2, 5\times10^{-5})	\nonumber \\
	& + 10^{3} \max(m_3, 5.48\times10^{-4}) \nonumber \\
	& + 10^{4} \max(m_4, 1.2\times10^{-4}) \nonumber \\
	& - 10^5 \text{GE} \nonumber \\
	& - 50 \text{ FTRWGE}
	\label{eq:tinytol_metric} 
\end{align}

When optimizing the 8-fold symmetric \toliman pupil, the same equation was used but with slightly different constants. Minor changes were found by experimentation to suit the new basis, resulting in the following objective function:


\begin{align}
	\mathscr{L}_\text{\toliman}  &= 10 \max(m_1, 0.0573) \nonumber \\
	& + 5\times10^{3} \max(m_2, 5\times10^{-5})	\nonumber \\
	& + 10^{3} \max(m_3, 5.48\times10^{-4}) \nonumber \\
	& + 2\times10^{4} \max(m_4, 1\times10^{-4}) \nonumber \\
	& - 2\times10^5 \text{GE} \nonumber \\
	& - 50 \text{ FTRWGE}
	\label{eq:toliman_metric} 
\end{align}

\subsubsection{Optimization}

\begin{figure}
    \centering
    \begin{tikzpicture}[scale = 1]
       
        \node at (-0.6, 1) [] {TinyTol Pupils};

        \node at (-5.25, 0.7) [] {Existing TinyTol Pupil};
    	\node[inner sep=0pt] () at (0, -2) {\includegraphics[height = 5cm]{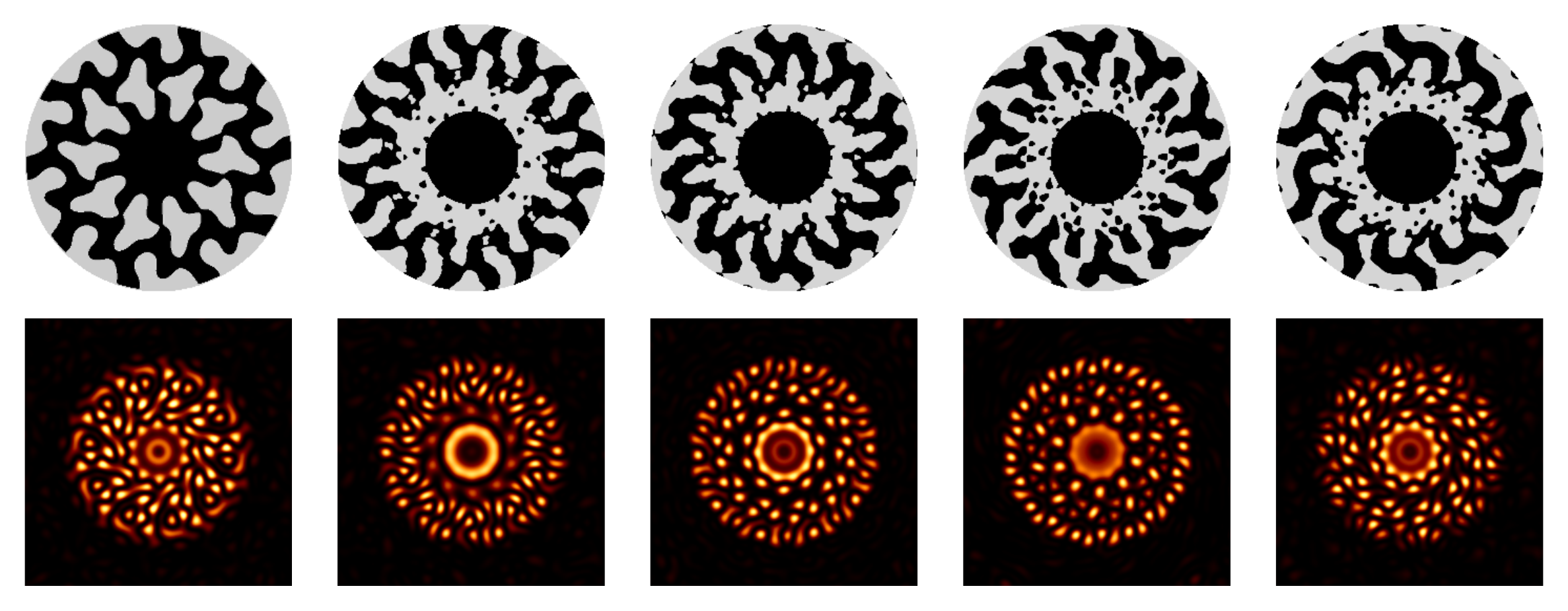}};
    	\node[inner sep=0pt] () at (7.05, 0) {\includegraphics[height = 1cm]{figures/tinytol/legend.pdf}};
        \node[inner sep=0pt] () at (6.65, -3.22) {\includegraphics[height = 2.37cm]{figures/gradient_desc/toliman_psf_zoom_cb.pdf}};
    	\draw (-6.5,-4.5) -- (-3.8,-4.5) -- (-3.8,0.4) -- (-6.5,0.4) -- (-6.5,-4.5);
    
        \node at (-0.6, -5.25) [] {\toliman Pupils};
        \node[inner sep=0pt] () at (0, -8) {\includegraphics[height = 5cm]{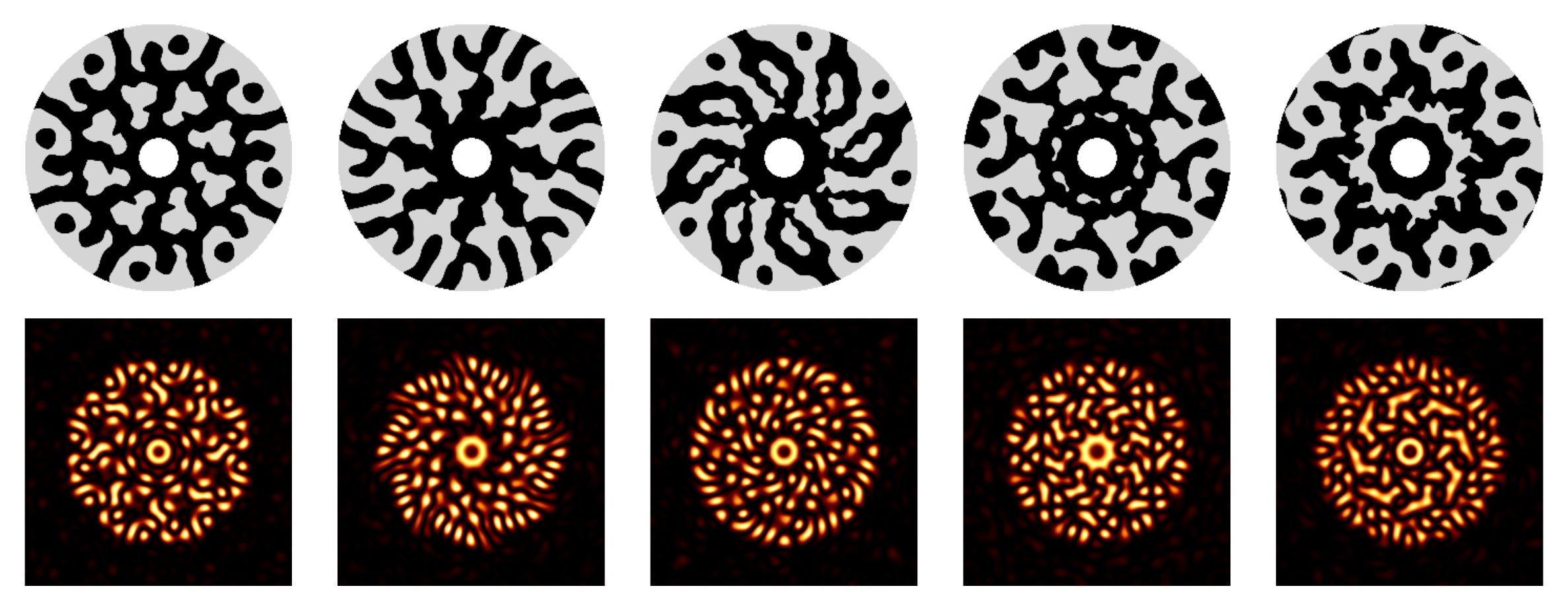}};
        \node[inner sep=0pt] () at (7.05, -6) {\includegraphics[height = 1cm]{figures/tinytol/legend.pdf}};
        \node[inner sep=0pt] () at (6.65, -9.22) {\includegraphics[height = 2.37cm]{figures/gradient_desc/toliman_psf_zoom_cb.pdf}};
    
    \end{tikzpicture}
    \caption{Sample diffractive pupils produced using the TinyTol (Equation~\ref{eq:tinytol_metric}) and \toliman (Equation~\ref{eq:toliman_metric}) metrics. Solution was initialized with a different seeds. Each solution achieves similar results on the objective function although the relative contributions of the different optimized terms vary. TinyTol color scales are normalized to the maximum value in the existing TinyTol pupil, while \toliman PSFs are normalized to the maximum across the ensemble.  \href{https://github.com/alipwong/phase_retrieval_and_design/tree/main/phase_design/Toliman}{\color{linkcolor}\faGithub}}
    \label{fig:toliman_pupils}
\end{figure}

Fitting for both TinyTol and \toliman was done by performing a gradient descent on the coefficients in the mode basis. 
We performed a simple fit under monochromatic conditions. Similar to the results in Figure~\ref{fig:qm_seeds}, the initialization affected the solution. We repeated gradient descent for multiple seeds (30 seeds for TinyTol, 50 seeds for \toliman). For TinyTol, we ran each fit for 250 epochs with a learning rate of $\eta$ = 0.05. For \toliman we ran for 500 epochs with a learning rate of $\eta$ = 0.05, which was dropped down to $\eta$ = 0.01 after 200 epochs. A selection of results are shown in Figure~\ref{fig:toliman_pupils}.

Computations were performed using the \textsc{Morphine} library\cite{pope2021}. It took $\sim$5.5 minutes to produce a TinyTol pupil (250 modes and 250 epochs) and $\sim$3 minutes to produce a \toliman pupil (80 modes and 500 epochs) on a MacBook Pro with~8~i9 cores and 32GB of RAM.


\subsubsection{Discussion and Results}


The solutions that arose from various seeds resulted in approximately the same overall merit, but their performance in individual metrics varied. 
Currently there is no global metric for choosing the optimum pupil, so these were chosen on visual appeal. Interestingly, the resultant family of solutions generally had similar structure --- large features at large radii and smaller features toward the center. 

We used a very simple gradient descent without an adaptive learning rate. During optimization, local gradients of the loss function sometimes exhibited sharp increases, which resulted in divergent solutions that never recovered. In these cases, the solution could either be discarded for solutions from other initializations, or they could be rerun with a lower learning rate.

\subsection{Practical Implementation}

This section has illustrated the potential to harness modern machine automatic differentiation algorithms to the hitherto challenging domain of phase design in optical systems.
The assumption at the outset is that masks only manipulate phase, and it is worth a brief digression here into what potential real-world implementations or methods of fabrication may be used to realize them. 
The short answer is that the machinery is quite general, and should prove of value wherever the pursuit of metrics quantifying desired outcomes is available.
Dynamically controllable examples include the deformable mirrors in AO systems, which furthermore can allow for the superimposition of the atmospheric correction while also adding a given phase mask pattern. 

Modern lithographic methods can etch static designs onto various transmissive or reflective optical surfaces that are either flat or curved. 
These systems provide spatial resolutions in the single-micron regime, providing realistic ways to fabricate very fine features and sharp structures.
Novel technologies like Multi-Twist Retarders \cite{komanduri2013} (used to fabricate the pupil in Figure~\ref{fig:true_toliman}) can be applied to transmissive optics with similar spatial resolutions, that provide nearly achromatic response through modifications in geometric phase.
As long as the phase mask is fit-for-purpose in its optical setting, the algorithms here can be used to formulate the design.

\section{Open Science}
\label{sec:open}

We have made the \textsc{Morphine} source code freely available on GitHub\footnote{\href{https://github.com/benjaminpope/morphine}{github.com/benjaminpope/morphine}}, available under a GPLv3 open-source license. Implementations of the key calculations are provided as Jupyter notebooks\footnote{\href{https://github.com/alipwong/phase_retrieval_and_design}{github.com/alipwong/phase\_retrieval\_and\_design}}, with GitHub links in the captions of each relevant figure. We encourage other researchers to reproduce, test, extend, and apply our work.
\section{Conclusions and further work}

We have taken the highly optimized automatic differentiation tools that underpin the modern deep learning revolution and applied them to optical simulation for astronomy. By using packages like \textsc{Jax} and \textsc{Tensorflow} to construct optical simulations and take derivatives, we have extended the work first done by Jurling \cite{jurling2014} in the realm of phase retrieval to more complex and constrained situations. We have shown that this permits optimization of very many parameters, including various modal bases and dense pixel grids, as well as through a neural network in order to perform global optimizations of the phases. With these tools the reconstruction of constrained pupil plane phases is simple, even in cases with information loss through non-flux conservative processes such as saturation, and with arbitrary and unusual PSFs.

The power of this method can be easily extended to complex imaging systems, where we have adapted the popular standard optical simulation package \textsc{Poppy} with autodiff capabilities, as \textsc{Morphine}. An advantage of this approach is that existing simulations in \textsc{Poppy} are easy to adapt to be differentiable, and take advantage of \textsc{Jax}'s just-in-time compilation, accelerated linear algebra, and GPU support. Using \textsc{Morphine} we showed promising results in astronomical phase mask design, specifically for the construction and optimization of an \ac{app} coronagraph. Basic gradient descent methods find solutions close to current state of the art. 

The gradient descent approach can even be used to optimize discrete phase maps, whether by regularization or through soft thresholding. In the latter case, the simple CLIMB algorithm proposed here allows simultaneous imposition of symmetry and binarization, and would extend well to shaped-pupil coronagraphs and other coded aperture optimization problems.

A limiting factor in the performance of current coronagraphs is their sensitivity to tip-tilt `jitter' and other low-order wavefront errors. It will be straightforward in our framework to optimize coronagraphs that are robust to these random errors by stochastic gradient descent over an ensemble of noise realizations. Similarly, by batching over wavelength diversity, we expect to be able to optimize broadband performance. By optimizing over phases in multiple conjugate planes, by analogy to multi-plane light conversion \cite{fontaine2018}, we expect that gradient descent approaches will enable the design of new and complex hybrid coronagraphs, imaging spectrographs, and astrophotonic devices.

There is no reason to limit future work to simply phase retrieval: any aspect of the imaging system that can be parametrized and differentiated can in principle be learned, calibrated, or optimized as part of the same approach. For example, the inter-pixel sensitivity flat field map could be jointly learned together with a PSF model (Desdoigts et al., in prep.).

\section*{Acknowledgments}

This research made use of NASA's Astrophysics Data System.

This research made use of \textsc{jax} \cite{jax}; \textsc{TensorFlow} \cite{tensorflow2015}; \textsc{Poppy}, an open-source optical propagation Python package originally developed for the James Webb Space Telescope project \cite{poppy}; the \textsc{IPython} package \cite{ipython}; \textsc{NumPy} \cite{numpy}; \textsc{matplotlib} \cite{matplotlib}; and \textsc{SciPy} \cite{scipy}.

BJSP is grateful to the School of Physics and Sydney Institute for Astronomy at the University of Sydney for hosting him as a visiting researcher during the COVID-19 pandemic. We acknowledge and pay respect to the Gadigal people of the Eora Nation. It is upon their unceded, sovereign, ancestral lands that the University of Sydney is built. BJSP would like to acknowledge the traditional owners of the land on which the University of Queensland is situated, the Turrbal and Jagera people. We pay respects to their Ancestors and descendants, who continue cultural and spiritual connections to Country.

\section*{Funding}

This research was supported by contributions from the Breakthrough Watch program which is managed by the Breakthrough Initiatives and sponsored by the Breakthrough Prize Foundation. 

This work was performed in part under contract with the Jet Propulsion Laboratory (JPL) funded by NASA through the Sagan Fellowship Program executed by the NASA Exoplanet Science Institute. 

\section*{Disclosures}

The authors declare no conflicts of interest.

\bibliography{ms}

\end{document}